\documentclass[a4paper,11pt]{article}
\usepackage{amsmath, amssymb, graphics, graphicx}
\usepackage{color}
\usepackage{physics}
\usepackage{longtable}
\usepackage{subfigure}
\usepackage{multirow}
\usepackage{slashed}
\usepackage{mathtools}
\usepackage{stackengine}
\stackMath

\newcommand{\mathsym}[1]{{}}

\newcommand{\rmd}{\mathrm{d}}
\newcommand{\rmi}{\mathrm{i}}

\newcommand{\SO}{\mathop{\rm SO}}

\allowdisplaybreaks

\usepackage[shortlabels]{enumitem}

\begin{document}


\begin{titlepage}
\begin{center}
\vspace*{-1.0cm}
\hfill {\footnotesize 06-01-2026}

\vspace{2.0cm}

{\LARGE  {\fontfamily{lmodern}\selectfont \bf Applied foliated conformal Carroll symmetries}} \\[.2cm]

\vskip 1.5cm
{\bf Eric A. Bergshoeff$^{\,a}$, Patrick Concha$^{b,c}$, Octavio Fierro$^{d}$, \\[.1truecm]
Evelyn Rodr\'\i guez$^{b,c}$ and Jan Rosseel$^e$}\\
\vskip 1.2cm

\begin{small}
{}$^a$ \textit{Van Swinderen Institute, University of Groningen \\
Nijenborgh 3, 9747 AG Groningen, The Netherlands} \\
\vspace{1mm}

\vspace{5mm}

{}$^b$ \textit{Departamento de Matem\'atica y F\'\i sica Aplicadas, \\
Universidad Cat\'olica de la Sant\'\i sima Concepci\'on,\\
Alonso de Ribera 2850, Concepci\'on, Chile} \\

\vspace{5mm}

{}$^c$ \textit{Grupo de Investigación en Física Teórica, GIFT, \\
	Universidad Cat\'olica de la Sant\'\i sima Concepci\'on,\\
Alonso de Ribera 2850, Concepci\'on, Chile} \\

\vspace{5mm}

{}$^d$ \textit{Facultad de Ingenier\'{i}a, Arquitectura y Diseño, Universidad San Sebasti\'{a}n,\\
Lientur 1457, Concepción 4080871, Chile}\\

\vspace{5mm}

{}$^e$ \textit{Division of Theoretical Physics, Rudjer Bo\v{s}kovi\'c Institute, \\
Bijeni\v{c}ka 54, 10000 Zagreb, Croatia} \\
\vspace{2mm}
{\texttt{e.a.bergshoeff[at]rug.nl, patrick.concha[at]ucsc.cl, ofierro27[at]yahoo.com, erodriguez[at]ucsc.cl,
Jan.Rosseel[at]irb.hr}}

\end{small}

\end{center}

\vskip .5cm
\begin{abstract}
\vskip.1cm\noindent

We apply the conformal compensating technique for constructing matter couplings to conformal scalars on a $D$-dimensional foliated conformal Carroll manifold dividing the tangent space into $(p+1)$-dimensional longitudinal and $(D-p-1)$-dimensional transversal directions corresponding to $p$-branes.  We show that the conformal Carroll algebra that was used for particle-like foliated geometries with $p=0$ cannot be used for higher-dimensional objects, called $p$-branes,  with $0 < p \le D-2$. Furthermore, string-like foliated geometries are not suitable for the conformal compensating technique due to the conformal invariance in the longitudinal directions that is present for $p=1$. All other cases can be dealt with provided one uses a different conformal extension of the Carroll algebra that amounts to a conformal extension in the longitudinal directions only supplemented with an additional an-isotropic dilatation.

By brane-duality similar results hold for foliated Galilean geometries which we present as well. Our results nicely  fit in with recent work on foliated Aristotelian geometries.

\end{abstract}

\end{titlepage}



%



\section{Introduction}

Carroll symmetries have been at the focus of several research  directions recently. They are the natural symmetries of any null surface including  black hole horizons \cite{Donnay:2019jiz} and the null infinity of flat spacetime. They even have occurred in recent investigations of black hole singularities \cite{Oling:2024vmq}. A
conformal extension of Carroll symmetries can be identified with the asymptotic BMS symmetries
of flat spacetime that play a fundamental role in Carroll holography \cite{Duval:2014uva,Bagchi:2023cen,Donnay:2023mrd}. Carroll gravity occurs in two different versions called electric Carroll gravity \cite{Henneaux:1979vn} and magnetic Carroll gravity \cite{Bergshoeff:2017btm}. A generalization of magnetic Carroll gravity to so-called $p$-brane magnetic Carroll gravity has been given in \cite{Bergshoeff:2023rkk}, with the theory of \cite{Bergshoeff:2017btm} corresponding to the case $p=0$. The $p$-brane Carroll gravity theories are based on $p$-brane Carroll geometries that are locally foliated in $(D-p-1)$-dimensional transversal leaves. Such foliated target space Carroll geometries have occurred in  recent investigations of de Sitter  holography \cite{Blair:2025nno} and Carroll limits of the AdS/CFT correspondence \cite{Fontanella:2025tbs}.

The conformal program is a useful way to construct matter-coupled gravity theories in a systematic way, see, e.g., \cite{Freedman:2012zz}. When applied to general relativity it amounts in its most basic form to establishing a
 relationship between the Einstein-Hilbert term in a second-order formulation  and the kinetic term of a real scalar field $\phi$ as follows:
\begin{equation}
\mathcal{L} \ \sim\ R\big(\Omega(E)\big)\hskip 1truecm  \leftrightarrow\hskip 1truecm  \mathcal{L}\ \sim\ E\phi\big[\Box + R(\Omega(E))\big]\phi\hskip 1truecm  \leftrightarrow\hskip 1truecm  \mathcal{L} \sim \phi\Box\phi\,.
\end{equation}
Here,  $E \equiv \det E_\mu{}^{\widehat A}$ and $\Omega(E)$ is the dependent Lorentz spin connection.\,\footnote{We use a hatted notation here since we wish to reserve the un-hatted index $A$ for later to indicate the longitudinal directions only.} To go from the Einstein-Hilbert term to the kinetic term for a scalar field, one first redefines the Vielbein as the scale-invariant product of a conformal Vielbein $E^{(C)\widehat A}_\mu$ and a compensating scalar $\phi$ as follows: $E_\mu{}^{\widehat A} = \phi E^{(C)\widehat A}_\mu{}$. In a second step one then restricts to a flat spacetime geometry $E^{(C)}_\mu{}^{\widehat A} = \delta_\mu{}^{\widehat A}$. Going in the opposite direction, one first  couples the real scalar field $\phi$ to conformal gravity obtained by `gauging' the conformal algebra and next  imposes the gauge-fixing condition $\phi =$ constant. Note that coupling a scalar to conformal gravity does not amount to replacing the derivatives in the Lagrangian by conformal covariant derivatives. This is related to the fact that the conformal gauge fields are not true connection fields. For instance, the dilatation gauge field $B_\mu$ which is absent in the above Lagrangian is not only the gauge field for dilatations but it also transforms with a shift under the special conformal transformations. The absence of $B_\mu$ in the Lagrangian  is then equivalent to the fact that the Lagrangian is invariant under these special conformal transformations. Moreover, the gauge field $F_\mu{}^{\hat A}$ for special conformal transformations is dependent and the expression for its trace explains the presence of the $R$ term in the above Lagrangian. Constructing a conformal gravitational coupling to a 
scalar is only the first step in the conformal program. To obtain a matter-coupled  gravity
theory one should for instance introduce $N$ scalars and use only one of them as a compensating scalar ending up with a  gravity theory coupled to $N - 1$ scalars.

Recently, we applied the conformal technique for $p=0$ to massless Carroll scalars and showed how the coupling of a Carroll scalar to conformal Carroll gravity, obtained by gauging  the conformally extended Carroll algebra, leads after gauge-fixing  to a Carroll gravity theory \cite{Bergshoeff:2024ilz}.\,\footnote{For a different approach to Carrollian matter couplings for $p=0$, see \cite{Afshar:2025imp}.}   There exist both electric Carroll scalars and magnetic Carroll scalars \cite{Henneaux:2021yzg,deBoer:2021jej} and we showed how each of them leads to a corresponding electric and magnetic Carroll gravity theory. A similar result, without using the compensating mechanism, was obtained in \cite{Baiguera:2022lsw}. Only magnetic Carroll gravity consists of curvature terms like in general relativity and has a second-order formulation. The electric Carroll gravity theory is quadratic in  so-called intrinsic torsion tensors without connection fields  and is more exotic. 

In the Carroll  case, we  worked with the particle Carroll limit of general relativity and we used the same particle Carroll limit of the conformal algebra. We will call the resulting algebra the `$0$-brane conformal Carroll algebra'. When extending the above approach from particles to $p$-branes with $p>0$ in the magnetic case, it seemed reasonable to expect that the  compensating mechanism would work again if we would now work with the $p$-brane Carroll limit of both general relativity and the conformal algebra. A bit to our surprise, we will show in this work that this does not work. Applying an iterative procedure one cannot couple a magnetic conformal Carroll scalar to Carroll conformal gravity for $p > 0$. In hindsight, the explanation for this obstruction  has to do with the fact that there is a discrete change in the $p$-brane Carroll limit of general relativity when going from  $p=0$ to $p>0$. 

In general, when taking the limit of the Einstein-Hilbert term, decomposing the Lorentz index $\hat A$ into a longitudinal index $A$ and a transversal index $a$ as $\hat A = (A,a)$ and rescaling the fields with a contraction parameter $\omega$, the Einstein-Hilbert term  decomposes into three  curvature terms: those  of longitudinal Lorentz rotations $R_{AB}(J^{AB})$, those of Carroll boosts  $R_{Aa}(J^{Aa})$ and those of  transversal rotations  $R_{ab}(J^{ab})$, together with two terms quadratic in the spin connections (see eq.~\eqref{eq:SEHcarexp}). The $R_{AB}(J^{AB})$ term is of leading power in the contraction parameter $\omega$ while the  other two curvature terms $R_{Aa}(J^{Aa})$ and $R_{ab}(J^{ab})$ and one of the spin connection terms have the same next-to-leading  power in $\omega$. Therefore, when taking the limit that $\omega \to \infty$ the resulting magnetic Carroll gravity theory is given by the curvature for longitudinal Lorentz rotations $R_{AB}(J^{AB})$. However, the particle case with $p=0$ is special since in that case both the curvature of longitudinal Lorentz rotations and the next-to-leading spin connection term vanish. This is a special case where the Carroll gravity Lagrangian is given  by the sum of the curvatures corresponding to  Carroll boosts and transversal rotations. A distinctive difference with the $p=0$ case is that the magnetic Carroll gravity Lagrangian for $p>0$  contains a single curvature term with longitudinal indices only. We will show in this work that this implies that, unlike the particle case with $p=0$, all the $p>0$ Carroll algebras need to be conformally extended in the longitudinal directions only except for an additional an-isotropic dilatation that acts in the transversal directions as well. 
The conformal technique does not work for the string case with $p=1$ due to the fact that the Carroll gravity Lagrangian for that case is already invariant under local scale transformations.

We will show via a formal brane duality \cite{Barducci:2018wuj,Bergshoeff:2020xhv} that similar results hold for Galilei gravity. Under this brane duality the longitudinal/transversal directions of the Carroll geometry are interchanged with the transversal/longitudinal  directions of the Galilei geometry. This duality shows that in the Galilei case, for $0\le p \le D-3$, the Galilei Lagrangian is proportional to the transversal Galilei curvature $R_{ab}(J^{ab})$ while for the domain wall case with $p=D-2$ the Lagrangian is proportional to a sum of the Galilean boost curvature $R_{Aa}(J^{Aa})$ and the longitudinal Galilean curvature $R_{AB}(J^{AB})$. Only in the domain wall case one should work with a limit of the relativistic conformal algebra which we denote as the `$p$-brane Galilean conformal algebra'. For all the other cases one should use a conformal extension of the  Galilei algebra in the transversal directions only except for an additional an-isotropic dilatation that acts in the longitudinal directions as well.
In the Galilean case, the  conformal technique does not apply for $p=D-3$, i.e.~defect-branes  with two transversal directions since for that case the corresponding Galilei gravity Lagrangian is already invariant under local dilatations.

This paper is organized as follows. In section 2 we give a  review of $p$-brane Carroll versus  Galilei gravity defined as $p$-brane limits of general relativity. We will focus on those aspects that are relevant for the present discussion.  In  subsection 3.1  we present the $p$-brane conformal Carroll algebra and show how it is related to the algebra of conformal isometries of a flat $p$-brane Carroll space-time. Next, in subsection 3.2 we construct the `gauging', i.e., a suitable set of dependent and independent gauge fields, of the $p$-brane conformal Carroll algebra.  In section 4 we show   that the consistent coupling of a Carroll scalar to these gauge fields relies on  an identity that is only valid for $p=0$  \cite{Bergshoeff:2024ilz}. This makes the coupling of this type of $p$-brane conformal Carroll gauge fields to a Carroll scalar problematic. In  section 5 we show that for all cases with $p>0$ the problem can be circumvented by working with a gauging of a different conformal  algebra. The only exception for which the conformal technique does not work is the string case due to the local scale symmetry of string Carroll gravity.  Finally, section 6 contains our conclusions and an outlook.
We have moved the discussion of the dual Galilean case to two separate Appendices. 
In Appendix A we discuss the conformal approach to the  case of general $p$-brane Galilei gravity except for the special domain wall case with  $p=D-2$. This special  case  requires a different conformal extension and is separately discussed in Appendix B.

\section{$p$-brane Carroll and $p$-brane Galilei gravity} \label{sec:pgravreview}

In this section we give a short review of $p$-brane Carroll and $p$-brane Galilei gravity defined as $p$-brane Carroll and Galilei limits of general relativity.  We will emphasize those aspects that are relevant for the remainder of this work. For more details, we refer to \cite{Bergshoeff:2023rkk}.

Both $p$-brane Carroll and $p$-brane Galilei geometry can be formulated as Cartan geometries with local structure group given by
\begin{align}
  \label{eq:structgroup}
  \left(\SO(1,p) \times \SO(D-p-1) \right) \ltimes \mathbb{R}^{(p+1)(D-p-1)} \,.
\end{align}
The $\SO(1,p)$ and $\SO(D-p-1)$ factors constitute what are called longitudinal Lorentz transformations and transversal rotations, whereas $\mathbb{R}^{(p+1)(D-p-1)}$ corresponds to $p$-brane Carrollian or $p$-brane Galilean boost transformations. The structure group \eqref{eq:structgroup} leaves two degenerate (co-)metrics invariant:
\begin{alignat}{4}
  \label{eq:flatmetrics}
  \text{for $p$-brane Carroll geometry} &: \qquad &  \tau^{\tilde{A}\tilde{B}} &= \left(\begin{matrix} \eta^{AB} & 0 \\ 0 & 0 \end{matrix}\right) \qquad & & \text{and} \qquad &  h_{\tilde{A}\tilde{B}} &= \left(\begin{matrix} 0 & 0 \\ 0 & \delta_{ab}\end{matrix}\right) \,, \nonumber \\
  \text{for $p$-brane Galilei geometry} &: \qquad &  \tau_{\tilde{A}\tilde{B}} &= \left(\begin{matrix} \eta_{AB} & 0 \\ 0 & 0 \end{matrix}\right) \qquad & & \text{and} \qquad &  h^{\tilde{A}\tilde{B}} &= \left(\begin{matrix} 0 & 0 \\ 0 & \delta^{ab}\end{matrix}\right) \,.
\end{alignat}
Here, the index $\tilde{A}$ takes on $D$ values: $\tilde{A} = 0, 1, \cdots, D-1$. It has been decomposed in indices $A$ and $a$. The index $A=0,1,\cdots,p$ is used to denote the fundamental representation of $\SO(1,p)$ and is called ``longitudinal'', while the index $a=p+1, p+2, \cdots, D-1$ is reserved for the fundamental representation of $\SO(D-p-1)$ and  is referred to as ``transversal''. In \eqref{eq:flatmetrics}, $\eta_{AB}$/$\eta^{AB}$ is a $(p+1)$-dimensional Minkowski metric/co-metric and will be used to freely raise and lower longitudinal indices in what follows. Likewise, $\delta_{ab}$/$\delta^{ab}$ is a $(D-p-1)$-dimensional Kronecker delta that is used to raise and lower transversal indices. In the Cartan formulation of $p$-brane Carroll and Galilei geometry, one introduces a solder (Vielbein) one-form, whose longitudinal and transversal components are denoted by $\tau_{\mu}{}^{A}$ and $e_{\mu}{}^{a}$. These transform as follows under the infinitesimal action of the structure group \eqref{eq:structgroup}:
\begin{subequations}
  \label{eq:deltatauecargal}
\begin{alignat}{2}
  \text{$p$-brane Carroll geometry} &: \qquad & \delta \tau_{\mu}{}^{A} &= -\lambda^{A}{}_{B} \tau_{\mu}{}^{B} - \lambda^{A}{}_{a} e_{\mu}{}^{a} \,, \nonumber \\
                                  & & \delta e_{\mu}{}^{a} &= -\lambda^{a}{}_{b} e_{\mu}{}^{b} \,. \label{eq:deltatauecar} \\
  \text{$p$-brane Galilei geometry} &: \qquad & \delta \tau_{\mu}{}^{A} &= -\lambda^{A}{}_{B} \tau_{\mu}{}^{B} \,, \nonumber \\
  & & \delta e_{\mu}{}^{a} &= -\lambda^{a}{}_{b} e_{\mu}{}^{b} - \lambda_{A}{}^{a} \tau_{\mu}{}^{A} \,. \label{eq:deltatauegal}                          
\end{alignat}
\end{subequations}
Here, $\lambda^{AB} = -\lambda^{BA}$ are the parameters of infinitesimal longitudinal Lorentz transformations, $\lambda^{ab} = -\lambda^{ba}$ those of infinitesimal transversal rotations and $\lambda^{Aa}$ those of infinitesimal $p$-brane Carrollian or Galilean boosts.

There is a formal duality under which $p$-brane Carroll geometry is exchanged with $(D-p-2)$-brane Galilei geometry \cite{Barducci:2018wuj,Bergshoeff:2020xhv}:
\begin{equation}\label{eq:duality1}
p-\textrm{brane\ Carroll geometry} \hskip .5truecm \longleftrightarrow\hskip .5truecm (D-p-2)-\textrm{brane\ Galilei\ geometry}\,.
\end{equation}
This duality is performed by re-interpreting the longitudinal/transversal Vielbeine $\tau_{\mu}{}^{A}$/$e_{\mu}{}^{a}$ of $p$-brane Carroll geometry as transversal/longitudinal Vielbeine $e_{\mu}{}^{b}$/$\tau_{\mu}{}^{B}$ of $(D-p-2)$-brane Galilei geometry:
\begin{equation} \label{eq:duality11}
(\tau_\mu{}^A\,, e_\mu{}^a)\hskip .5truecm \longleftrightarrow \hskip .5truecm (e_\mu{}^b\,, \tau_\mu{}^B)\,,
\end{equation}
where range $A$ = range $b$ and range $a$ = range $B$. To stay in line with the convention that the time direction is always longitudinal, one has to supplement this formal interchange with a Wick rotation to change the signatures of the degenerate flat metrics \eqref{eq:flatmetrics} accordingly, i.e., one has to replace $\eta_{A A^\prime}$ by $\delta_{b b^\prime}$ and $\delta_{a a^\prime}$ by $\eta_{B B^\prime}$. For the convenience of the reader, we will in the following treat the $p$-brane Carroll and $p$-brane Galilei cases separately, even though they are formally related via the above duality. Physically, they correspond to very different (non-relativistic vs. ultra-relativistic) limits of Lorentzian geometry. We first discuss the Carroll case and next the Galilei case.

As we already mentioned in the introduction there are two types of Carroll gravity theories, called electric and magnetic Carroll gravity, that have  rather different properties. An electric $p$-brane Carroll gravity theory is obtained by taking a $p$-brane Carroll limit of the Einstein-Hilbert (EH) action in a second-order formulation. A noteworthy feature of this limit is that the resulting action is quadratic in so-called intrinsic torsion tensors and thus does not contain any spin connection field. Therefore, after taking the limit one obtains a second-order formulation that no longer admits a standard first-order Cartan formulation.\,\footnote{A non-standard first-order formulation of $0$-brane electric Carroll gravity, in which the fields transform in a different representation than that obtained from gauging the Carroll algebra, was presented in \cite{Pekar:2024ukc}.} These electric Carroll gravity theories are therefore rather unconventional. In contrast, a magnetic Carroll gravity theory can be obtained by taking a $p$-brane Carroll limit of the EH action in a first-order formulation. The resulting action does contain spin connection fields and does allow a second-order formulation. They are  more in line with a conventional gravity theory. Below we provide some more details on magnetic $p$-brane Carroll gravity and its Galilean equivalent.

\vskip .2truecm

\noindent {\bf magnetic $p$-brane Carroll gravity}\ \
\vskip .1truecm

To derive $p$-brane Carroll gravity, we start from the $D$-dimensional EH action
\begin{align}
  \label{eq:EH}
S_{\rm EH} & = - \frac{1}{16 \pi G_{\rm N}} \int \rmd^D x \, E E_{\hat{A}}{}^\mu E_{\hat{B}}{}^\nu{} R_{\mu\nu}{}^{\hat{A}\hat{B}}(\Omega) 
\end{align}
in the first-order formulation. Here, $G_{\rm N}$ is Newton's constant, $E_\mu{}^{\hat A}\ (\mu,\hat A = 0,1,\cdots,D-1)$ is the relativistic Vielbein and $E_{\hat{A}}{}^{\mu}$ is its inverse:
\begin{equation}
E_\mu{}^{\hat A} E_{\hat{B}}{}^\mu = \delta^{\hat A}_{\hat B}\,,\hskip 2truecm E_\mu{}^{\hat A} E_{\hat{B}}{}^\nu = \delta_\mu^\nu\,.
\end{equation}
Furthermore, $E = \mathrm{det}(E_\mu{}^{\hat{A}})$ denotes the determinant of the Vielbein and
the curvature $R_{\mu\nu}{}^{\hat{A}\hat{B}}(\Omega)$ is defined in terms of the relativistic spin connection field $\Omega_\mu{}^{\hat A\hat B}$ as follows: 
\begin{equation}
R_{\mu\nu}{}^{\hat{A}\hat{B}}(\Omega) = 2\partial_{[\mu}\Omega_{\nu]}{}^{\hat A\hat B} + 2 \Omega_{[\mu}{}^{\hat A\hat C} \Omega_{\nu]}{}_{\hat C}{}^{\hat B} \,.
\end{equation}

To define the $p$-brane Carroll limit of general relativity we first decompose the index $\hat A$ into a longitudinal index $A\ (=0,1,\cdots ,p)$ and a transversal index $a\ (= p+1, \cdots ,D-1)$. We next redefine the Vielbeine, spin connection fields and Newton's constant with a dimensionless contraction parameter $\omega$  as follows:
\begin{equation}
    \begin{aligned}\label{eq:redefCarr}
        E_\mu{}^A &= \omega^{-1}\, \tau_\mu{}^A \,, \,\hspace{.75cm} &&E_\mu{}^a = e_\mu{}^a   \,, \hspace{.75cm} &&G_{\rm N} = \omega^{\alpha} G_{\rm C}\,,\\
        \Omega_\mu{}^{AB} \;&=\; \omega_\mu{}^{AB} \,, &&\Omega_\mu{}^{ab} \;=\; \omega_\mu{}^{ab}\,, \hspace{.75cm} &&\Omega_\mu{}^{Aa} \;=\; \omega^{-1} \omega_\mu{}^{Aa} \,,
    \end{aligned}
\end{equation}
where $\alpha \in \mathbb{R}$ will be determined soon. After using the redefinitions \eqref{eq:redefCarr} in the EH action \eqref{eq:EH}, one finds that the latter can be expanded in powers of $\omega$ as:
\begin{align}
  \label{eq:SEHcarexp}
  S_{\rm EH} &= - \frac{\omega^{-(p+1) - \alpha}}{16 \pi G_{\rm C}} \int \rmd^D x \, e \Big[ \omega^2 \tau_{A}{}^{\mu} \tau_{B}{}^{\nu} R_{\mu\nu}(J^{AB}) + e_{a}{}^{\mu} e_{b}{}^{\nu} R_{\mu\nu}(J^{ab}) + 2 \tau_{A}{}^{\mu} e_{a}{}^{\nu} R_{\mu\nu}(G^{Aa}) \nonumber \\ & \qquad \qquad \qquad \qquad \qquad - 2 \tau_{A}{}^{\mu} \tau_{B}{}^{\nu} \omega_{[\mu}{}^{Aa} \omega_{\nu]}{}^{B}{}_{a} - 2 \omega^{-2} e_{a}{}^{\mu} e_{b}{}^{\nu} \omega_{[\mu}{}^{Aa} \omega_{\nu]A}{}^{b} \Big] \,.
\end{align}
Here, $e=\det(\tau_\mu{}^A\,, e_\mu{}^a)$ and the different curvatures are defined by
\begin{align}\label{eq:curvatures}
  R_{\mu\nu}(J^{AB}) &= 2\,\partial_{[\mu}\omega_{\nu]}{}^{AB} + 2\omega_{[\mu}{}^{AC}\omega_{\nu]C}{}^B \,, \nonumber \\[.1truecm]
  R_{\mu\nu}(J^{ab}) &= 2\partial_{[\mu}\omega_{\nu]}{}^{ab} + 2\omega_{[\mu}{}^{ac}\omega_{\nu]c}{}^b\,, \nonumber \\[.1truecm]
R_{\mu\nu}(G^{Aa}) &=  2\,\partial_{[\mu}\omega_{\nu]}{}^{Aa} + 2\omega_{[\mu}{}^{AB}\omega_{\nu]B}{}^a + 2\omega_{[\mu}{}^{ab}\omega_{\nu]}{}^A{}_b\,.
\end{align}
The Carroll limit of $S_{\mathrm{EH}}$ corresponds to the $\omega \to \infty$ limit of \eqref{eq:SEHcarexp}, i.e., to retaining only the terms that are of leading order in $\omega$. When $p \neq 0$, this amounts to keeping the first term of \eqref{eq:SEHcarexp} that involves the curvature $R_{\mu\nu}(J^{AB})$ of the spin connection for the longitudinal Lorentz transformations. In the special case that $p=0$ there are no longitudinal Lorentz transformations. The first and fourth terms of \eqref{eq:SEHcarexp} are then absent and the terms that are of leading order in $\omega$ are given by the second and third ones. Summarizing, we find the following Carroll actions
\begin{align}
\text{for $p=0$}:\hskip .5truecm  &S_{\rm Carroll} =  -\frac{1}{16\pi G_{\rm C}} \int \rmd^D x \, e\ \Big[ 2\tau_A{}^\mu e_a{}^\nu R_{\mu\nu}(G^{Aa}) + e_a{}^\mu e_b{}^\nu R_{\mu\nu}(J^{ab})\Big ]\,,\label{eq:special}\\[.1truecm]
\text{for $0<p<D-1$}: \hskip .5truecm &S_{\rm Carroll} = -\frac{1}{16\pi G_{\rm C}} \int \rmd^D x \, e\ \tau_A{}^\mu \tau_B{}^\nu   R_{\mu\nu}(J^{AB})\,,\label{eq:generic}
\end{align}
where we have taken
\begin{equation}
\text{for $p=0$}:\ \alpha = - 1 \,,\hskip 2truecm  \text{for $0 < p < D-1$}:\ \alpha = -(p-1)\,,
\end{equation}
to ensure that the terms in \eqref{eq:SEHcarexp} that are of leading order in $\omega$ are the $\mathcal{O}(\omega^0)$ ones. 

Going to a second-order formulation, not all spin connection components in the above actions can be solved for. Some components occur as Lagrange multipliers and therefore remain independent. Depending on $p$ these independent spin connection components are given by
\begin{align}
p=0:\hskip .5truecm &\omega_{(a|,0|b)}\,,\\[.1truecm]
p=1:\hskip .5truecm &\omega_{a,AB}\ \ {\rm and}\ \ \omega_{B,AB}\,,\label{carrp1}\\[.1truecm]
1<p<D-1:\hskip .5truecm &\omega_{a,AB}\,,
\end{align}
where $0$ indicates the flat time direction.

We note that the generic Carroll gravity actions \eqref{eq:generic} with  $0 < p <D -1$ are invariant under the following emergent local an-isotropic scale transformations with parameter $\lambda_{D}$:
\begin{equation}
\delta \tau_\mu{}^A = (D-p-1)\lambda_{D} \tau_\mu{}^A\,,\hskip 1truecm \delta e_\mu{}^a = -(p-1)\lambda_D\, e_\mu{}^a\,.\nonumber
\end{equation}

The above shows that there are two particular cases that turn out to play a special role when considering a conformal approach to $p$-brane Carroll gravity,
\vskip .2truecm

\noindent $p=0:$ All Carroll actions given in eqs.~\eqref{eq:special} and \eqref{eq:generic} are invariant under Carroll boost transformations but only  the particle Carroll gravity action \eqref{eq:special} contains the Carroll boost spin connection in a non-trivial way.
\vskip .1truecm

\noindent $p=1:$ Only the $p=1$ ``string'' Carroll action with two longitudinal directions is invariant under a local an-isotropic scale transformation that does not act on the transversal Vielbein fields. Furthermore, all components of the longitudinal spin connection remain independent, see eq.\,\eqref{carrp1}.

We next turn our attention to  the Galilei case.
\vskip .3truecm

\noindent {\bf magnetic $p$-brane Galilei Gravity}\ \
\vskip .1truecm

We again start from the first-order $D$-dimensional EH action \eqref{eq:EH}. To define the $p$-brane Galilei limit, we proceed similarly as in the Carroll case. We begin by making the following redefinitions involving a dimensionless parameter $\omega$:
\begin{equation}
    \begin{aligned}\label{eq:redefGal}
        E_\mu{}^A &= \omega\, \tau_\mu{}^A \,, \hspace{.75cm} &&E_\mu{}^a = e_\mu{}^a \,, \hspace{.75cm} &&G_{\rm N} = \omega^{\alpha} G_{\rm G} \,, \\
        \Omega_\mu{}^{AB} \;&=\; \omega_\mu{}^{AB} \,, &&\Omega_\mu{}^{ab} \;=\; \omega_\mu{}^{ab}\,, \hspace{.75cm} &&\Omega_\mu{}^{Aa} \;=\; \omega^{-1} \omega_\mu{}^{Aa} \,,
    \end{aligned}
\end{equation}
for real $\alpha$. The Galilei limit of the EH action is obtained by taking the $\omega \rightarrow \infty$ limit of \eqref{eq:EH}, i.e., by using the redefinitions \eqref{eq:redefGal} in \eqref{eq:EH} and keeping only the terms that dominate as $\omega$ tends to infinity. For $0 \leq p < D - 2$, only one term that involves the curvature $R_{\mu\nu}(J^{ab})$ of the spin connection of transversal rotations is kept in this limit. Similar to what happened for the Carroll limit, there is one special case, namely $p= D - 2$, when there are no transversal rotations. The result for the limit of \eqref{eq:EH} then consists of two terms that involve the curvatures $R_{\mu\nu}(J^{AB})$ and $R_{\mu\nu}(G^{Aa})$. Concretely, the $\omega \rightarrow \infty$ limit of the EH action leads to the following Galilei actions:
\begin{align}
\text{for $p=D-2$}:\hskip .5truecm  &S_{\rm Galilei} =  -\frac{1}{16\pi G_{\rm G}} \int \rmd^D x \, e\ \Big[
  \tau_A{}^\mu \tau_B{}^\nu R_{\mu\nu}(J^{AB}) +
 2\tau_A{}^\mu e_a{}^\nu R_{\mu\nu}(G^{Aa})\Big ]\,,\label{specialG}\\[.1truecm]
\text{for $0\le p<D-2$}: \hskip .5truecm &S_{\rm Galilei} = -\frac{1}{16\pi G_{\rm G}} \int \rmd^D x \, e\ e_a{}^\mu e_b{}^\nu   R_{\mu\nu}(J^{ab})\,,\label{genericG}
\end{align}
where we have taken
\begin{equation}
\text{for $p=D-2$}:\ \alpha = D-3 \,,\hskip 2truecm  \text{for $0<p<D-1$}:\ \alpha = p+1\,,
\end{equation}
to make sure that the $\mathcal{O}(\omega^0)$ terms are the ones that are dominant in the $\omega \rightarrow \infty$ limit.

Going to a second-order formulation, the following spin connection components occur as Lagrange multipliers in the above actions and therefore remain independent;
\begin{align}\
p=D-2:\hskip .5truecm &\omega_{(A|,z|B)}\,,\label{gn1}\\[.1truecm]
p=D-3:\hskip .5truecm &\omega_{A,ab}\ \ {\rm and}\ \ \omega_{b,ab}\,,\label{gn2}\\[.1truecm]
0\le p <D-3:\hskip .5truecm &\omega_{A,ab}\,,\label{gn3}
\end{align}
where $z$ is the transversal direction of the domain wall.

We note that the generic Galilei gravity actions \eqref{genericG} with  $0 \le p <D -2$ are invariant under the following emergent local an-isotropic scale transformations with parameter $\lambda_D$:
\begin{equation}
\delta \tau_\mu{}^A = -(D-p-3)\lambda_{D} \tau_\mu{}^A\,,\hskip 1truecm \delta e_\mu{}^a = (p+1)\lambda_D\, e_\mu{}^a\,.
\end{equation}

The above shows that there are two particular cases that turn out to  play a special role  when considering a conformal approach to $p$-brane Galilei gravity,
\vskip .2truecm

\noindent $p=D-2$: All Galilei actions given in eqs.~\eqref{specialG} and \eqref{genericG} are invariant under Galilei boost transformations but only the `domain wall' Galilei  gravity action \eqref{specialG} contains the Galilei boost spin connection $\omega_{\mu}{}^{Aa}$ in a non-trivial way.
\vskip .1truecm

\noindent $p=D-3$: Only the `defect brane' Galilei action with two transversal directions is invariant under a local an-isotropic scale transformation that does not act on the longitudinal Vielbein fields. In this case, all components of the transverse spin connection remain independent, see eq.\,\eqref{gn2}.

This finishes our short review of the $p$-brane Carroll and $p$-brane Galilei actions.

\section{The $p$-brane conformal Carroll algebra and its gauging} \label{sec:pconfcarrgauging}

In this section, we will consider the gauging of the $p$-brane conformal Carroll algebra. This algebra is presented in subsection \ref{ssec:pconfcarralg}, where we also comment on how it is related to the algebra of conformal isometries of a flat $p$-brane Carroll space-time. Subsection \ref{ssec:pconfcarrgauging} is devoted to the details of the gauging of the $p$-brane conformal Carroll algebra. The results of this section will be used in section \ref{sec:confcalccarrp}, where we will investigate whether they can serve as a starting point to reconstruct $p$-brane Carroll gravity actions via conformal calculus methods.

\subsection{The $p$-brane conformal Carroll algebra} \label{ssec:pconfcarralg}

The $p$-brane conformal Carroll algebra, that we will denote by $\mathfrak{ccarr}_{p}$ in what follows, can be obtained as an In\"on\"u-Wigner contraction of the relativistic conformal algebra $\mathfrak{so}(2,D)$. The algebra $\mathfrak{ccarr}_{p}$ is spanned by the following generators:
\begin{align}
  \label{eq:pconfcarrgens}
  \mathfrak{ccarr}_{p} &= \mathrm{Span}\{H_{A}, P_{a}, J_{AB}, J_{ab}, G_{Aa}, L_{A}, K_{a}, D\} \,.
\end{align}
$H_{A}$ and $P_{a}$ generate translations in the longitudinal and transversal directions and will accordingly be referred to as longitudinal and transversal translation generators. Likewise, $J_{AB}$ and $J_{ab}$ span the Lie algebras $\mathfrak{so}(1,p)$ and $\mathfrak{so}(D-p-1)$ of Lorentz transformations of the longitudinal directions and rotations of the transversal ones and will be said to generate longitudinal Lorentz transformations and transversal rotations. $G_{Aa}$ on the other hand will be called generators of $p$-brane Carroll boosts (or boosts for short), since they boost longitudinal directions into transversal ones but not vice versa. Together $H_{A}$, $P_{a}$, $J_{AB}$, $J_{ab}$ and $G_{Aa}$ constitute the $p$-brane Carroll algebra. The extra generators that extend the $p$-brane Carroll algebra to $\mathfrak{ccarr}_{p}$ consist of the generators $L_{A}$ of longitudinal special conformal transformations, the generators $K_{a}$ of transversal special conformal transformations, as well as a dilatation generator $D$. The non-zero commutation relations of $\mathfrak{ccarr}_{p}$ are given by:
\begin{alignat}{3}
  \label{eq:pconfcarr}
  \commutator{J_{AB}}{J_{CD}} &= 4 \eta_{[A[C} J_{D]B]} \,, \qquad \ \ & \commutator{J_{ab}}{J_{cd}} &= 4 \delta_{[a[c} J_{d]b]} \,, \qquad \ \ & 
  \commutator{G_{Aa}}{J_{BC}} &= 2 \eta_{A[B} G_{C]a} \,, \nonumber \\ \commutator{G_{Aa}}{J_{bc}} &= 2 \delta_{a[b|} G_{A|c]} \,, \qquad \ \ &
  \commutator{H_{A}}{J_{BC}} &= 2 \eta_{A[B} H_{C]} \,, \qquad \ \ & \commutator{P_{a}}{J_{bc}} &= 2 \delta_{a[b} P_{c]} \,, \nonumber \\
  \commutator{L_{A}}{J_{BC}} &= 2 \eta_{A[B} L_{C]} \,, \qquad \ \ & \commutator{K_{a}}{J_{bc}} &= 2 \delta_{a[b} K_{c]} \,, \qquad \ \ &
  \commutator{P_{a}}{G_{Ab}} &= - \delta_{ab} H_{A} \,, \nonumber \\ \commutator{K_{a}}{G_{Ab}} &= - \delta_{ab} L_{A} \,, \qquad \ \ & 
  \commutator{D}{H_{A}} &= H_{A} \,, \qquad \ \ & \commutator{D}{P_{a}} &= P_{a} \,, \nonumber \\
  \commutator{D}{L_{A}} &= - L_{A} \,, \qquad \ \ & \commutator{D}{K_{a}} &= - K_{a} \,, \qquad \ \ &
  \commutator{H_{A}}{K_{a}} &= 2 G_{Aa} \,, \nonumber \\  \commutator{P_{a}}{L_{A}} &= -2 G_{Aa} \,, \qquad \ \ &
  \commutator{P_{a}}{K_{b}} &= 2 \delta_{ab} D + 2 J_{ab} \,.
\end{alignat}
From these commutation relations, one sees that the generators $\{D, L_A, K_a, J_{AB}, J_{ab}, G_{Aa}\}$ span a subalgebra that we will denote by $\mathfrak{hccarr}_p$ and that we will call the subalgebra of homogeneous $p$-brane conformal Carroll transformations. The commutators of $H_A$, $P_a$, $L_A$ and $K_a$ with $D$ furthermore show that the latter generates isotropic dilatations.

In order to clarify the physical interpretation of the $p$-brane conformal Carroll transformations, we consider a flat $p$-brane Carroll space-time. In an inertial set of coordinates $x^{\mu} = (x^{A}, x^{a})$, the longitudinal co-metric $\tau = \tau^{\mu\nu} \partial_{\mu} \otimes \partial_{\nu}$ of rank-$(p+1)$ and the transversal metric $h = h_{\mu\nu} \rmd x^{\mu} \otimes \rmd x^{\nu}$ of rank-$(D-p-1)$ are given by
\begin{align}
  \label{eq:flatcarrmetric}
  \tau = \eta^{AB} \partial_{A} \otimes \partial_{B} \,, \qquad \qquad h = \delta_{ab} \rmd x^{a} \otimes \rmd x^{b} \,.
\end{align}
The infinitesimal action of the $p$-brane conformal Carroll transformations on the $(x^{A}, x^{a})$ coordinates can be found by taking a $p$-brane Carroll limit of the infinitesimal action of the relativistic conformal transformations. On Minkowski coordinates $x^{\mu}$, the latter is given by:
\begin{align}
  \label{eq:relconf}
  \delta x^{\mu} &= a^{\mu} + m^{\mu}{}_{\nu} x^{\nu} + \lambda x^{\mu} + 2 \kappa_{\nu} x^{\nu} x^{\mu} - \kappa^{\mu} x^{\nu} x_{\nu} \,.
\end{align}
Here, $a^{\mu}$, $m_{\mu\nu} = - m_{\nu\mu}$, $\lambda$ and $\kappa^{\mu}$ are the (constant) parameters of infinitesimal translations, Lorentz transformations, dilatations and special conformal transformations. To take the $p$-brane Carroll limit, we first split the index $\mu$ in $A$ and $a$ and rescale the coordinates and symmetry parameters with a dimensionless parameter $\omega$ as follows:
\begin{alignat}{4}
  x^{A} \ &\rightarrow \ \omega^{-1}\, x^{A} \,, \qquad & x^{a} \ &\rightarrow \ x^{a} \,, \qquad & a^{A} \ &\rightarrow \ \omega^{-1}\, a^{A} \,, \qquad & a^{a} \ &\rightarrow \ a^{a} \,, \nonumber \\
  m_{AB} \ &\rightarrow \ m_{AB} \,, \qquad & m_{Aa} \ &\rightarrow \ \omega^{-1}\, m_{Aa} \,, \qquad & m_{ab} \ &\rightarrow \ m_{ab} \,, \qquad & \lambda \ &\rightarrow \ \lambda \,, \nonumber \\
  \kappa_{A} \ &\rightarrow \ \omega^{-1}\, \kappa_{A} \,, \qquad & \kappa_{a} \ &\rightarrow \ \kappa_{a} \,.
\end{alignat}
Using these redefinitions in \eqref{eq:relconf}, we find that the $\omega \rightarrow \infty$ limit is well-defined. Taking this limit leads to the infinitesimal action of the $p$-brane conformal Carroll transformations on the $(x^{A}, x^{a})$ coordinates:
\begin{align}
  \label{eq:confcarrtrafos}
  \delta x^{A} &= a^{A} + m^{A}{}_{B} x^{B} + m^{A}{}_{a} x^{a} + \lambda x^{A} + 2 \kappa_{a} x^{a} x^{A} - \kappa^{A} x^{a} x_{a} \,, \nonumber \\
  \delta x^{a} &= a^{a} + m^{a}{}_{b} x^{b} + \lambda x^{a} + 2 \kappa_{b} x^{b} x^{a} - \kappa^{a} x^{b} x_{b}  \,.
\end{align}
The parameters $a^{A}$, $a^{a}$, $m^{AB}$, $m^{ab}$, $m^{Aa}$, $\lambda$, $\kappa^{A}$ and $\kappa^{a}$ are now interpreted as those of the transformations generated by $H_{A}$, $P_{a}$, $J_{AB}$, $J_{ab}$, $G_{Aa}$, $D$, $L_{A}$ and $K_{a}$.

The $p$-brane conformal Carroll transformations \eqref{eq:confcarrtrafos} should be contrasted with the $p$-brane conformal Carroll isometries, i.e., the conformal isometries of a flat $p$-brane Carroll space-time. The conformal isometries of a generic $p$-brane Carroll space-time with (co-)metrics $\tau^{\mu\nu}$ and $h_{\mu\nu}$ are generated by those infinitesimal general coordinate transformations $\delta x^{\mu} = \xi^{\mu}$ that satisfy the following ``$p$-brane conformal Carroll Killing equations'' for some function $\Lambda$:
\begin{subequations}
  \label{eq:carrconfiso}
  \begin{alignat}{3}
    \mathcal{L}_\xi \tau^{\mu\nu} &= - \Lambda \tau^{\mu\nu} \qquad \quad & & \Leftrightarrow \qquad \quad & & \xi^{\rho} \partial_{\rho} \tau^{\mu\nu} - 2 \partial_{\rho} \xi^{(\mu} \tau^{\nu)\rho} = - \Lambda \tau^{\mu\nu} \,, \label{eq:carrconfisotau} \\
    \mathcal{L}_\xi h_{\mu\nu} &= \Lambda h_{\mu\nu} \qquad \quad & & \Leftrightarrow \qquad \quad & & \xi^{\rho} \partial_{\rho} h_{\mu\nu} + 2 \partial_{(\mu} \xi^{\rho} h_{\nu)\rho} = \Lambda h_{\mu\nu} \,. \label{eq:carrconfisoh}
  \end{alignat}
\end{subequations}
For the flat $p$-brane Carroll metric structure \eqref{eq:flatcarrmetric}, these equations take the following form:
\begin{subequations}
  \label{eq:confkilleqs}
  \begin{align}
    & \partial_{A} \xi^{a} = 0 \,, \label{eq:confkilleqs1} \\
    & \partial_{A} \xi_{B} + \partial_{B} \xi_{A} = \Lambda \eta_{AB} \,, \label{eq:confkilleqs2} \\
    & \partial_{a} \xi_{b} + \partial_{b} \xi_{a} = \Lambda \delta_{ab} \,. \label{eq:confkilleqs3} 
  \end{align}
\end{subequations}
For $p \neq D-2$ or $p \neq D-3$,\footnote{For $p=D-2$, there is only one transversal coordinate that we denote by $y$, so we split $x^{\mu} = (x^{A}, y)$ and $\xi^{\mu} = (\xi^{A}, \xi^{y})$. The equations \eqref{eq:confkilleqs} for $p=D-2$ imply that $\xi^{y} = \xi^{y}(y)$ is an arbitrary function of $y$ and that $\xi^{A} = a^{A}(y) + m^{A}{}_{B}(y) x^{B} + \partial_y \xi^{y} x^{A}$, with $a^{A}(y)$ and $m_{AB}(y) = - m_{BA}(y)$ arbitrary functions of $y$. For $p=D-3$, there are two transversal directions, whose coordinates we denote by $y^1$ and $y^2$, so we split $x^{\mu} = (x^{A}, y^1, y^2)$ and $\xi^{\mu} = (\xi^{A}, \xi^{y^1}, \xi^{y^2})$. We can then introduce complex coordinates $z = y^1 + \rmi y^2$, $\bar{z} = y^1 - \rmi y^2$ and complex combinations $\xi = \xi^{y^1} + \rmi \xi^{y^2}$, $\bar{\xi} = \xi^{y^1} - \rmi \xi^{y^2}$. The equations \eqref{eq:confkilleqs} for $p=D-3$ then imply that $\xi = \xi(z)$ is holomorphic (and that $\bar{\xi} = \bar{\xi}(\bar{z})$ is anti-holomorphic) and that $\xi^{A} = a^{A}(z,\bar{z}) + m^{A}{}_{B}(z,\bar{z}) x^{B} + \frac12 \left(\partial_z \xi + \partial_{\bar{z}} \bar{\xi}\right) x^{A}$, where $a^{A}(z,\bar{z})$ and $m_{AB}(z,\bar{z}) = - m_{BA}(z,\bar{z})$ are arbitrary real functions of $z$, $\bar{z}$.} the generic solution of these equations leads to the following infinitesimal action of the $p$-brane conformal Carroll isometries on the coordinates $(x^{A}, x^{a})$:
\begin{alignat}{2}
  \label{eq:Xconfisos}
  \delta x^{A} &= \xi^{A} & &= a^{A}(x^{a}) + m^{A}{}_{B}(x^{a}) x^{B} + \lambda x^{A} + 2 \kappa_{a} x^{a} x^{A} \,, \nonumber \\
  \delta x^{a} &= \xi^{a} & &= a^{a} + m^{a}{}_{b} x^{b} + \lambda x^{a} + 2 \kappa_{b} x^{b} x^{a} - \kappa^{a} x^{b} x_{b} \,,
\end{alignat}
where $a^{a}$, $m_{ab} = - m_{ba}$, $\lambda$, $\kappa^{a}$ are constant and $a^{A}(x^{a})$ and $m_{AB}(x^{a}) = - m_{BA}(x^{a})$ are arbitrary functions of the transversal coordinates $x^{a}$. Comparing \eqref{eq:confcarrtrafos} with \eqref{eq:Xconfisos}, one sees that the actions of the $p$-brane conformal Carroll transformations and that of the $p$-brane conformal Carroll isometries agree on the transversal $x^{a}$ coordinates, but not on the longitudinal $x^{A}$ coordinates. In particular, the translation and longitudinal Lorentz transformation parameters $a_{A}$ and $m_{AB}$ get promoted to functions $a_{A}(x^{a})$ and $m_{AB}(x^{a})$ in the action of the $p$-brane conformal Carroll isometries. The boosts $m^{A}{}_{a} x^a$ and longitudinal special conformal transformations $\kappa^{A} x^{a} x_{a}$ of \eqref{eq:confcarrtrafos} are then just special cases of the conformal Carroll isometries with parameters $a^{A}(x^{a})$. From the above discussion, we thus conclude that the infinitesimal $p$-brane conformal Carroll transformations constitute a finite-dimensional subalgebra of the infinite-dimensional algebra of the infinitesimal $p$-brane conformal Carroll isometries.

This concludes our review of the $p$-brane conformal Carroll algebra. In the next subsection, we will discuss its gauging.

\subsection{Gauging the $p$-brane conformal Carroll algebra} \label{ssec:pconfcarrgauging}

In this subsection, we will gauge the algebra $\mathfrak{ccarr}_p$. The results of this subsection will be used in section \ref{sec:confcalccarrp} to determine to which extent this gauging can be used to construct conformally invariant actions for a compensating scalar that are gauge equivalent to $p$-brane Carroll gravity actions in the second-order formulation. 

In the first step of the gauging procedure, we introduce the $\mathfrak{ccarr}_p$-valued gauge field
\begin{align}
  \label{eq:A}
  A_\mu &= \tau_\mu{}^A H_A + e_\mu{}^a P_a + \frac12 \omega_{\mu}{}^{AB} J_{AB} + \omega_{\mu}{}^{Aa} G_{Aa} + \frac12 \omega_{\mu}{}^{ab} J_{ab} + f_{\mu}{}^{A} L_A + g_{\mu}{}^{a} K_a + b_\mu D \,,
\end{align}
and gauge parameter
\begin{align}
  \label{eq:Sigma}
  \Sigma &= \zeta^A H_A + \zeta^a P_a + \frac12 \lambda^{AB} J_{AB} + \lambda^{Aa} G_{Aa} + \frac12 \lambda^{ab} J_{ab} + \lambda_K{}^A L_A + \lambda_K{}^a K_a + \lambda_D D \,.
\end{align}
The gauge transformation rules of the component fields $\tau_{\mu}{}^{A}$, $e_{\mu}{}^{a}$, $\omega_{\mu}{}^{AB}$, $\omega_{\mu}{}^{Aa}$, $\omega_{\mu}{}^{ab}$, $f_{\mu}{}^{A}$, $g_{\mu}{}^{a}$ and $b_{\mu}$ can be extracted from the general gauge transformation rule of $A_{\mu}$:
\begin{align}
  \label{eq:deltaA}
  \delta A_\mu &= \partial_\mu \Sigma + \commutator{A_{\mu}}{\Sigma} \,.
\end{align}
In what follows, we will mostly ignore the local longitudinal and transversal translations with parameters $\zeta^{A}$, $\zeta^{a}$ and focus on the homogeneous $p$-brane conformal Carroll transformations instead. Denoting the infinitesimal action of the latter with the symbol $\delta_0$, we find from \eqref{eq:deltaA} that
\begin{subequations}
  \label{eq:hccarrptrafos}
\begin{align}
  \delta_0 \tau_{\mu}{}^{A} &= - \lambda^{A}{}_{B} \tau_{\mu}{}^{B} - \lambda^{Aa} e_{\mu a} - \lambda_D \tau_{\mu}{}^{A} \,, \qquad \qquad \delta_0 \tau_{A}{}^{\mu} = - \lambda_{A}{}^{B} \tau_{B}{}^{\mu} + \lambda_D \tau_{A}{}^{\mu} \,, \label{eq:d0tau} \\
    \delta_0 e_{\mu}{}^{a} &= - \lambda^{a}{}_{b} e_{\mu}{}^{b} - \lambda_D e_{\mu}{}^{a} \,, \qquad \qquad \qquad \qquad \quad \, \delta_0 e_{a}{}^{\mu} = - \lambda_{a}{}^{b} e_{b}{}^{\mu} + \lambda^{A}{}_{a} \tau_{A}{}^{\mu} + \lambda_D e_{a}{}^{\mu} \,, \label{eq:d0e} \\
  \delta_0 \omega_{\mu}{}^{AB} &= \partial_{\mu} \lambda^{AB} - 2 \lambda^{[A|C} \omega_{\mu C}{}^{|B]} \,, \label{eq:d0omAB} \\
  \delta_0 \omega_{\mu}{}^{Aa} &= \partial_{\mu} \lambda^{Aa} + \omega_{\mu}{}^{A}{}_{B} \lambda^{Ba} + \omega_{\mu}{}^{a}{}_{b} \lambda^{Ab} - \lambda^{A}{}_{B} \omega_{\mu}{}^{Ba} - \lambda^{a}{}_{b} \omega_{\mu}{}^{Ab} - 2 \lambda_K{}^{A} e_{\mu}{}^a  \nonumber \\ & \quad + 2 \lambda_K{}^a \tau_{\mu}{}^{A} \,, \label{eq:d0omAa} \\
  \delta_0 \omega_{\mu}{}^{ab} &= \partial_{\mu} \lambda^{ab} - 2 \lambda^{[a|c} \omega_{\mu c}{}^{|b]} - 4 \lambda_K{}^{[a} e_{\mu}{}^{b]} \,, \label{eq:d0omab} \\
  \delta_0 f_{\mu}{}^{A} &= \partial_{\mu} \lambda_K{}^{A} + \omega_{\mu}{}^{A}{}_{B} \lambda_K{}^{B} + \omega_{\mu}{}^{Aa} \lambda_{K a} - \lambda^{A}{}_{B} f_{\mu}{}^{B} - \lambda^{Aa} g_{\mu a} - \lambda_K{}^A b_{\mu} \nonumber \\ & \quad + \lambda_D f_{\mu}{}^{A} \,, \label{eq:d0f} \\
  \delta_0 g_{\mu}{}^{a} &= \partial_{\mu} \lambda_K{}^{a} + \omega_{\mu}{}^{a}{}_{b} \lambda_K{}^{b} - \lambda^{a}{}_{b} g_{\mu}{}^{b} - \lambda_K{}^{a} b_{\mu} + \lambda_D g_{\mu}{}^{a} \,, \label{eq:d0g} \\
  \delta_0 b_{\mu} &= \partial_{\mu} \lambda_D + 2 \lambda_K{}^a e_{\mu a} \,. \label{eq:d0b}
\end{align}
\end{subequations}
The field strengths that are covariant with respect to these homogeneous conformal Carroll transformations are defined as follows:
\begin{subequations}
  \label{eq:ccarrpcurvs}
\begin{align}
  R_{\mu\nu}(H^{A}) &= 2 \partial_{[\mu} \tau_{\nu]}{}^{A} + 2 \omega_{[\mu}{}^{AB} \tau_{\nu]B} + 2 \omega_{[\mu}{}^{Aa} e_{\nu]a} + 2 b_{[\mu} \tau_{\nu]}{}^A \,, \label{eq:RH} \\
  R_{\mu\nu}(P^{a}) &= 2 \partial_{[\mu} e_{\nu]}{}^{a} + 2 \omega_{[\mu}{}^{ab} e_{\nu]b} + 2 b_{[\mu} e_{\nu]}{}^{a} \,, \label{eq:RP} \\
  R_{\mu\nu}(J^{AB}) &= 2 \partial_{[\mu} \omega_{\nu]}{}^{AB} + 2 \omega_{[\mu}{}^{[A|C} \omega_{\nu]C}{}^{|B]} \,, \label{eq:RJAB} \\
  R_{\mu\nu}(J^{ab}) &= 2 \partial_{[\mu} \omega_{\nu]}{}^{ab} + 2 \omega_{[\mu}{}^{[a|c} \omega_{\nu]c}{}^{|b]} + 8 g_{[\mu}{}^{[a} e_{\nu]}{}^{b]} \,, \label{eq:RJab} \\
  R_{\mu\nu}(G^{Aa}) &= 2 \partial_{[\mu} \omega_{\nu]}{}^{Aa} + 2 \omega_{[\mu}{}^{AB} \omega_{\nu]B}{}^{a} + 2 \omega_{[\mu}{}^{ab} \omega_{\nu]}{}^{A}{}_{b} - 4 g_{[\mu}{}^{a} \tau_{\nu]}{}^{A} + 4 f_{[\mu}{}^{A} e_{\nu]}{}^a \,, \label{eq:RGAa} \\
  R_{\mu\nu}(L^{A}) &= 2 \partial_{[\mu} f_{\nu]}{}^{A} + 2 \omega_{[\mu}{}^{AB} f_{\nu]B} + 2 \omega_{[\mu}{}^{Aa} g_{\nu]a} + 2 f_{[\mu}{}^{A} b_{\nu]} \,, \label{eq:RL} \\
  R_{\mu\nu}(K^{a}) &= 2 \partial_{[\mu} g_{\nu]}{}^{a} + 2 \omega_{[\mu}{}^{ab} g_{\nu]b} + 2 g_{[\mu}{}^{a} b_{\nu]} \,, \label{eq:RK} \\
  R_{\mu\nu}(D) &= 2 \partial_{[\mu} b_{\nu]} - 4 g_{[\mu}{}^{a} e_{\nu]a} \,.
\end{align}
\end{subequations}

Thus far, the gauging of $\mathfrak{ccarr}_p$ leads to a multiplet of \emph{independent} fields $\tau_{\mu}{}^{A}$, $e_{\mu}{}^{a}$, $\omega_{\mu}{}^{AB}$, $\omega_{\mu}{}^{Aa}$, $\omega_{\mu}{}^{ab}$, $f_{\mu}{}^{A}$, $g_{\mu}{}^{a}$, $b_{\mu}$ that transform under $\mathfrak{hccarr}_p$, according to the rules \eqref{eq:hccarrptrafos}. Starting from this multiplet, one can construct an $\mathfrak{hcarr}_{p}$-multiplet with less independent fields. This is done by imposing so-called conventional constraints on components of the curvatures \eqref{eq:ccarrpcurvs} that contain components of the dilatation gauge field, the spin connections $\omega_{\mu}{}^{AB}$, $\omega_{\mu}{}^{Aa}$, $\omega_{\mu}{}^{ab}$ and special conformal gauge fields $f_{\mu}{}^{A}$, $g_{\mu}{}^{a}$ algebraically and linearly. These dilatation, spin connection and special conformal gauge field components can then be viewed as \emph{dependent} fields, since they can be expressed in terms of the remaining gauge field components by solving them from the conventional constraints.

A maximal set of conventional constraints, that leads to a multiplet with the minimal amount of independent fields, is given by:
\begin{alignat}{3}
  \label{eq:convconstr}
  R_{Aa}(P^a) &= 0 \,, \quad & R_{A[a}(P_{b]}) &= 0 \,, \qquad & R_{bc}(P^a) &= 0 \,, \nonumber \\
  R_{BC}(H^A) &= 0 \,, \qquad & R_{Ba}(H^A) &= 0 \,, \qquad & R_{ab}(H^A) &= 0 \,, \nonumber \\
  R_{\mu b}(J^{ab}) &= 0 \,, \qquad & R_{\mu a}(G^{Aa}) &= 0 \,.
\end{alignat}
The first six of these constraints can be used to express the components
\begin{align}
  \label{eq:depbom}
  b_A\,, \quad \omega_{A, ab} \,, \quad \omega_{c, ab} \,, \quad \omega_{C, AB} \,, \quad \omega_{B, Aa} \,, \quad \omega_{[a|,A|b]} \,,
\end{align}
in terms of the remaining independent field components:
\begin{align}
  \label{eq:indepfields}
 \tau_{\mu}{}^{A} \,, \quad e_{\mu}{}^{a} \,, \quad b_a \,, \quad \omega_{a, AB} \,, \quad \omega_{(a|,A|b)} \,.
\end{align}
In particular, the expressions for the components \eqref{eq:depbom} that are obtained from the first six constraints of \eqref{eq:convconstr} are given by
\begin{alignat}{2}
  \label{eq:depbomsols}
  \omega_{A, ab} &= e_{A[a,b]} \,, \qquad \qquad & \omega_{c,ab} &= e_{c[a,b]} - \frac12 e_{ab,c} - 2 b_{[a} \delta_{b]c} \,, \nonumber \\
  \omega_{B,Aa} &= -\tau_{Ba,A} + \omega_{a,AB} + b_a \eta_{AB} \,, \qquad \qquad & \omega_{[a|,A|b]} &= - \frac12 \tau_{ab, A} \,, \nonumber \\
  \omega_{B,Aa} &= -\tau_{Ba,A} + \omega_{a,AB} + b_a \eta_{AB} \,, \qquad \qquad & b_A &= - \frac{1}{D-p-1} e_{Aa}{}^a \,.
\end{alignat}
Here, we have used the notation:
\begin{align}
  \tau_{\mu\nu}{}^{A} \equiv 2 \partial_{[\mu} \tau_{\nu]}{}^{A} \,, \qquad \qquad \qquad e_{\mu\nu}{}^{a} \equiv 2 \partial_{[\mu} e_{\nu]}{}^a \,.
\end{align}
From the last two conventional constraints of \eqref{eq:convconstr}, one can solve for the special conformal gauge fields $f_{\mu}{}^{A}$ and $g_{\mu}{}^{a}$:
\begin{align}
  \label{eq:depfgsols}
  f_{\mu}{}^{A} &= - \frac{1}{2(D-p-2)} \Big[ R^\prime_{\mu a}(G^{Aa}) - \frac{1}{(D-p-1)} \tau_{\mu}{}^{B} R^\prime_{Ba}(G^{Aa}) - \frac{1}{2(D-p-1)} \tau_{\mu}{}^{A} R^\prime_{ab}(J^{ab}) \Big] \,, \nonumber \\
  g_{\mu}{}^{a} &= -\frac{1}{2(D-p-3)} \Big[ R^\prime_{\mu b}(J^{ab}) - \frac{1}{2(D-p-2)} e_{\mu}{}^{a} R^\prime_{bc}(J^{bc}) - \frac{1}{(D-p-2)} \tau_{\mu}{}^{A} R^\prime_{Ab}(J^{ab}) \Big] \,,
\end{align}
where
\begin{align}
  \label{eq:Rprime}
  R_{\mu\nu}^\prime(G^{Aa}) &= 2 \partial_{[\mu} \omega_{\nu]}{}^{Aa} + 2 \omega_{[\mu}{}^{AB} \omega_{\nu]B}{}^{a} + 2 \omega_{[\mu}{}^{ab} \omega_{\nu]}{}^{A}{}_{b} \,, \nonumber \\
  R_{\mu\nu}^\prime(J^{ab}) &= 2 \partial_{[\mu} \omega_{\nu]}{}^{ab} + 2 \omega_{[\mu}{}^{[a|c} \omega_{\nu]c}{}^{|b]} \,.
\end{align}
From now on, it will be understood that the dependent components \eqref{eq:depbom}, as well as $f_{\mu}{}^{A}$ and $g_{\mu}{}^{A}$, are given by their explicit expressions \eqref{eq:depbomsols} and \eqref{eq:depfgsols} in terms of the independent fields \eqref{eq:indepfields}.

The conventional constraints \eqref{eq:convconstr} thus lead to a minimal realization of the homogeneous $p$-brane conformal Carroll algebra on the independent fields \eqref{eq:indepfields} that constitute a proper subset of all components of the original $\mathfrak{ccarr}_p$-valued gauge field \eqref{eq:A}. In what follows, we will denote the action of $\mathfrak{hccarr}_p$ transformations in this minimal realization by the symbol $\delta$. On the independent fields \eqref{eq:indepfields}, these $\delta$ transformations agree with the $\delta_0$ ones of \eqref{eq:hccarrptrafos}, i.e., one has:
\begin{align}
  \label{eq:hccarrpindep}
  \delta \tau_{\mu}{}^{A} &= - \lambda^{A}{}_{B} \tau_{\mu}{}^{B} - \lambda^{Aa} e_{\mu a} - \lambda_D \tau_{\mu}{}^{A} \,, \nonumber \\
  \delta e_{\mu}{}^{a} &= - \lambda^{a}{}_{b} e_{\mu}{}^{b} - \lambda_D e_{\mu}{}^{a} \,, \nonumber \\
  \delta \tau_{A}{}^{\mu} &= - \lambda_{A}{}^{B} \tau_{B}{}^{\mu} + \lambda_D \tau_{A}{}^{\mu} \,, \nonumber \\
  \delta e_{a}{}^{\mu} &= - \lambda_{a}{}^{b} e_{b}{}^{\mu} + \lambda^{A}{}_{a} \tau_{A}{}^{\mu} + \lambda_D e_{a}{}^{\mu} \,, \nonumber \\
  \delta b_{a} &= \partial_{a} \lambda_{D} + \lambda_{D} b_{a} - \lambda_{a}{}^{b} b_{b} + \lambda^{A}{}_{a} b_{A} + 2 \lambda_{K a} \,, \nonumber \\
  \delta \omega_{a}{}^{AB} &= \partial_{a} \lambda^{AB} - 2 \lambda^{[A|C} \omega_{a, C}{}^{|B]} - \lambda_{a}{}^{b} \omega_{b}{}^{AB} + \lambda^{C}{}_{a} \omega_{C}{}^{AB} + \lambda_{D} \omega_{a}{}^{AB} \,, \nonumber \\
  \delta \omega_{(a|,A|b)} &= \partial_{(a} \lambda_{|A|b)} + \lambda^{B}{}_{(a} \omega_{b), AB} + \lambda^{B}{}_{(a} \omega_{|B,A|b)} + \lambda_{A}{}^{c} \omega_{(a,b)c} - \lambda_{A}{}^{B} \omega_{(a|,B|b)} \nonumber \\ & \quad - \lambda_{a}{}^{c} \omega_{(b|,A|c)} - \lambda_{b}{}^{c} \omega_{(a|,A|c)} + \lambda_D \omega_{(a|,A|b)} - 2 \lambda_{K A} \delta_{ab} \,.
\end{align}

The transformation rules of the dependent field components \eqref{eq:depbom} in the minimal realization are found by varying the expressions \eqref{eq:depbomsols} and \eqref{eq:depfgsols} under \eqref{eq:hccarrpindep}. The resulting transformations do not necessarily coincide with the $\delta_0$ rules (that can be found from \eqref{eq:hccarrptrafos}) for the corresponding field components, but might instead include extra terms that we denote by $\Delta$. The $\delta$ transformations in the minimal realization can thus in general be written as
\begin{align}
  \delta = \delta_0 + \Delta \,,
\end{align}
where $\Delta$ is zero when acting on the independent fields \eqref{eq:indepfields}. The extra $\Delta$ terms on the dependent fields are needed to ensure that the conventional constraints \eqref{eq:convconstr} are invariant under the full $\delta$ transformations. Explicitly, one finds that the extra $\Delta$ contributions to the transformation rules of the dependent fields \eqref{eq:depbom} are given by the following boost transformations:
\begin{subequations}
  \label{eq:Deltacontribs}
\begin{align}
  \Delta b_{A} &= \frac{1}{(D-p-1)} \lambda^{B}{}_{a} e_{BA}{}^{a} \,, \label{eq:DbA} \\
  \Delta \omega_{A,ab} &= \lambda^{B}{}_{[a} e_{|AB|,b]} \,, \label{eq:DomAab} \\
  \Delta \omega_{c,ab} &= - \lambda^{A}{}_{a} e_{A\{b,c\}} + \lambda^{A}{}_{b} e_{A\{a,c\}} \,, \label{eq:Domcab} \\
  \Delta \omega_{C,AB} &= - \lambda_{[A}{}^{a} e_{B]C,a} + \frac12 \lambda_{C}{}^{a} e_{AB,a} - \frac{2}{(D-p-1)} \lambda^{Da} e_{D[A|,a} \eta_{|B]C} \,, \label{eq:DomCAB} \\
  \Delta \omega_{B}{}^{A}{}_{a} &= \lambda^{Ab} e_{B\{a,b\}} \,, \label{eq:DomBAa} \\
  \Delta \omega_{[a}{}^{A}{}_{b]} &= 0 \,. \label{eq:DomaAbasym} 
\end{align}
\end{subequations}
The transformation rules of the special conformal gauge fields $f_{\mu}{}^{A}$ and $g_{\mu}{}^{a}$ also acquire extra infinitesimal boosts $\Delta f_{\mu}{}^{A}$ and $\Delta g_{\mu}{}^{a}$. We only give these $\Delta$ transformations for the traces $f_{A}{}^{A}$ and $g_{a}{}^{a}$, since these are the only components that are needed to construct conformally invariant actions for scalar fields in the conformal approach. We can give $\Delta g_{a}{}^{a}$ and $\Delta f_{A}{}^{A}$ in terms of the $\Delta$ variations \eqref{eq:Deltacontribs}: 
\begin{subequations}
  \label{eq:Deltagftraces}
\begin{align}
  \Delta g_{a}{}^{a}  &= -\frac{1}{2(D-p-2)} D_{a}^{\prime} \Delta \omega_{b}{}^{ab} + \frac{1}{2(D-p-2)} \omega_{a}{}^{A}{}_{b} \Delta \omega_{A}{}^{ab} \,, \label{eq:Deltagaaoms} \\
  \Delta f_{A}{}^{A} &= \frac{1}{2(D-p-1)} \Bigg(D_{a}^\prime \Delta \omega_{A}{}^{Aa} + \frac{(p+1)}{(D-p-2)} D_{a}^\prime \Delta \omega_{b}{}^{ab} + \omega_{A,Ba} \Delta \omega^{B,Aa} - \omega_{a}{}^{Ba} \Delta \omega_{A}{}^{A}{}_{B} \nonumber \\ & \qquad \qquad \qquad + \omega_{A}{}^{Ab} \Delta \omega_{a}{}^{a}{}_{b} - \frac{(D-1)}{(D-p-2)} \omega_{a,Ab} \Delta \omega^{A,ab} - \delta_0 R_{Aa}(G^{Aa}) \Bigg) \,. \label{eq:DeltafAAoms}
\end{align}
\end{subequations}
Here, the derivative $D_{\mu}^{\prime}$ is covariantized with respect to $\omega_{\mu}{}^{AB}$, $\omega_{\mu}{}^{ab}$, $b_{\mu}$, but not with respect to $\omega_{\mu}{}^{Aa}$, e.g., one has:
\begin{align}
  D_{a}{}^{\prime} \Delta \omega_{b}{}^{ab} &\equiv \partial_{a} \Delta \omega_{b}{}^{ab} + \omega_{a}{}^{ac} \Delta \omega_{bc}{}^{b} - b_{a} \Delta \omega_{b}{}^{ab} \,.
\end{align}
The result \eqref{eq:DeltafAAoms} for $\Delta f_{A}{}^{A}$ also contains the $\delta_0$ variation of the boost curvature trace $R_{Aa}(G^{Aa})$. This is explicitly given by:
\begin{align}
  \label{eq:d0RAaGAa}
  \delta_0 R_{Aa}(G^{Aa}) &= \lambda^{B}{}_{a} R_{AB}(G^{Aa}) + \lambda^{Ba} R_{Aa}(J^{A}{}_{B}) \,.  
\end{align}
For $\Delta f_{A}{}^{A} + \Delta g_{a}{}^{a}$, one then finds
\begin{align}
\label{eq:Deltafgtraces}
  \Delta f_{A}{}^{A} + \Delta g_{a}{}^{a} &= \frac{1}{2(D-p-1)} \Bigg(D_{a}^{\prime} \Delta \omega_{A}{}^{Aa} - \frac{(D-2p-2)}{(D-p-2)} D_{a}^{\prime} \Delta \omega_{b}{}^{ab} + \omega_{A,Ba} \Delta \omega^{B,Aa} \nonumber \\ & \qquad \qquad - \omega_{a}{}^{Ba} \Delta \omega_{A}{}^{A}{}_{B} + \omega_{A}{}^{Ab} \Delta \omega_{a}{}^{a}{}_{b} - \frac{p}{(D-p-2)} \omega_{a,Ab} \Delta \omega^{A,ab} \nonumber \\ & \qquad \qquad - \delta_0 R_{Aa}(G^{Aa}) \Bigg) \,.
\end{align}
Restricting to the $p=0$ case, one sees that all terms on the right-hand-side are either zero or cancel. This cancellation was crucial in the construction of 0-brane magnetic Carroll gravity via the conformal approach \cite{Bergshoeff:2024ilz}. For non-zero $p$, such a simplification does not seem to take place. In particular, focusing on the first two terms on the right-hand-side (i.e., the only terms that involve derivatives of the boost parameter), one finds using \eqref{eq:DomBAa} and \eqref{eq:Domcab} that
\begin{align}
  \frac{1}{2(D-p-1)} \left(D_{a}^\prime \Delta \omega_{A}{}^{Aa} - \frac{(D-2p-2)}{(D-p-2)} D_{a}^{\prime} \Delta \omega_{b}{}^{ab} \right) \nonumber \\ \qquad \qquad = \frac{p}{2(D-p-1)(D-p-2)} D_{a}^{\prime} \left(\lambda^{A}{}_{b} e_{A}{}^{\{a,b\}} \right) \,.
\end{align}
These terms thus do not vanish when $p \neq 0$. In the next section, we will attempt to reconstruct $p$-brane Carroll gravity from a conformal method based on the above results for the gauging of the $p$-brane conformal Carroll algebra. We will see that the non-vanishing of $\Delta f_{A}{}^{A} + \Delta g_{a}{}^{a}$ forms an obstruction to obtain $p$-brane magnetic Carroll gravity in this attempt.

\section{Probing a conformal Approach to $p$-brane Carroll Gravity}\label{sec:confcalccarrp}

In our earlier work \cite{Bergshoeff:2024ilz}, the gauging of the $0$-brane conformal Carroll algebra was used to reconstruct $0$-brane electric and magnetic Carroll gravity from the conformal approach. In this procedure, one first renders the actions of a $0$-brane electric and magnetic Carroll scalar invariant under local $0$-brane conformal Carroll transformations, by coupling them to the independent and dependent gauge fields that are obtained from the gauging. It was then shown in \cite{Bergshoeff:2024ilz} that the resulting actions are gauge equivalent to those of $0$-brane electric and magnetic Carroll gravity. Here, we will use the results for the gauging of the $p$-brane conformal Carroll algebra of section \ref{sec:pconfcarrgauging} to check whether the approach of \cite{Bergshoeff:2024ilz} can be generalized to non-zero values of $p$. For $0$-brane magnetic Carroll gravity, the construction of \cite{Bergshoeff:2024ilz} crucially relied on the fact that the particular combination \eqref{eq:Deltafgtraces} of the $\Delta$ boost transformation of the dependent special conformal gauge fields vanishes. As we saw below \eqref{eq:Deltafgtraces}, this property is specific for $p=0$ and an analogous statement no longer holds when $p \neq 0$. The conclusion of this section is then that it is not straightforward to use the gauging of the $p$-brane conformal Carroll algebra to construct actions that are gauge equivalent to the $p$-brane magnetic Carroll gravity ones. For this reason, we will explore a conformal calculus based on a different algebra in the next section \ref{sec:pconfargrav}.

We start from the Lagrangians of the $p$-brane analogues of an electric and a magnetic Carroll scalar in flat space-time (with the metric structure \eqref{eq:flatcarrmetric}). For a $p$-brane electric scalar $\phi$, this Lagrangian is given by:
\begin{align}
  \label{eq:Lflatelscal}
  \mathcal{L}_{\text{el.}} &= -\frac12 \partial^{A} \phi \partial_{A} \phi \,. 
\end{align}
The Lagrangian of $p$-brane magnetic scalar $\phi$ on the other hand is in flat space-time given by
\begin{align}
  \label{eq:Lflatmagnscal}
  \mathcal{L}_{\text{magn.}} &= \pi^{A} \partial_{A} \phi - \frac12 \partial^{a} \phi \partial_{a} \phi \,,
\end{align}
where $\pi^{A}$ is a longitudinal vector that acts as a Lagrange multiplier for the constraint $\partial_{A} \phi = 0$. 

In line with what was done in \cite{Bergshoeff:2024ilz}, we then wish to make the actions \eqref{eq:Lflatelscal} and \eqref{eq:Lflatmagnscal} invariant under local homogeneous $p$-brane conformal Carroll transformations. To do this, we assume that $\phi$ transforms under dilatations with a scaling weight $w$:
\begin{align}
  \label{eq:dilscal}
  \delta \phi &= w \lambda_D \phi \,,
\end{align}
and that it is inert under the other $p$-brane homogeneous conformal Carroll transformations. We then define the following dilatation-covariant transversal and longitudinal derivatives:
\begin{align}
  \label{eq:covderphi}
  D_{A} \phi &= \tau_{A}{}^{\mu} \left(\partial_{\mu} \phi - w b_{\mu} \phi \right) \,, \qquad \qquad \qquad D_{a} \phi = e_{a}{}^{\mu} \left(\partial_{\mu} \phi - w b_{\mu} \phi \right) \,.
\end{align}
Using \eqref{eq:d0b} and \eqref{eq:DbA}, these are found to transform as follows under homogeneous $p$-brane conformal Carroll transformations:
\begin{align}
  \label{eq:trafosDphi}
  \delta D_{A} \phi &= -\lambda_{A}{}^{B} D_{B} \phi + (w+1) \lambda_D D_{A} \phi - \frac{w}{D-p-1} \lambda^{B}{}_{a} e_{BA}{}^{a} \phi \,, \nonumber \\
  \delta D_{a} \phi &= -\lambda_{a}{}^{b} D_{b} \phi + \lambda^{A}{}_{a} D_{A} \phi + (w+1) \lambda_D D_{a} \phi - 2 w \lambda_{Ka} \phi \,.
\end{align}

\subsection{Electric $p$-brane Carroll Gravity}

Let us consider the electric Lagrangian \eqref{eq:Lflatelscal} first. As an initial try to make this invariant under local homogeneous $p$-brane conformal Carroll transformations, we replace the $\partial_{A}$ derivatives by covariant $D_{A}$ ones:
\begin{align}
  \label{eq:L0elconf}
  \mathcal{L}_{0, \text{el. conf.}} &= - \frac{e}{2} D^{A} \phi D_{A} \phi \,,
\end{align}
where $e = \det(\tau_{\mu}{}^{A}, e_{\mu}{}^{a})$. This is invariant under local dilatations, provided the scaling weight $w$ is chosen as
\begin{align}
  \label{eq:w}
  w = \frac{D-2}{2} \,.
\end{align}
For $p\neq 0$, the Lagrangian \eqref{eq:L0elconf} is however not invariant under $p$-brane Carrollian boosts, due to the fact that the dependent gauge field $b_{A}$ acquires the extra boost transformation $\Delta b_{A}$ given in \eqref{eq:DbA}. In particular, one finds the following boost variation of \eqref{eq:L0elconf}:
\begin{align}
  \label{eq:deltaL0elconf}
  \delta \mathcal{L}_{0, \text{el. conf.}} &= e \frac{w}{(D-p-1)} \lambda^{B}{}_{a} e_{BA}{}^{a} \phi D^{A}\phi \,.
\end{align}
This variation can be cancelled by introducing an extra field $\pi_{AB}{}^{a} = \pi_{[AB]}{}^{a}$ that transforms under homogeneous $p$-brane conformal Carroll transformations as follows:
\begin{align}
  \label{eq:piABatrafo}
  \delta \pi_{AB}{}^{a} &= -2 \lambda_{[A}{}^{C} \pi_{|C|B]}{}^{a} - \lambda^{a}{}_{b} \pi_{AB}{}^{b} - \frac{w}{2(D-p-1)} \lambda_{[A}{}^{a} D_{B]}\left(\phi^2\right) + \left(2 w + 1\right) \lambda_D \pi_{AB}{}^{a} \,.
\end{align}
The following Lagrangian
\begin{align}
  \mathcal{L}_{\text{el. conf.}} &= - \frac{e}{2} D^{A} \phi D_{A} \phi + e \pi^{AB}{}_{a} e_{AB}{}^{a} \,,
\end{align}
is then invariant under local homogeneous $p$-brane conformal Carroll transformations, provided the scaling weight $w$ is chosen as in \eqref{eq:w}. Fixing the dilatations by adopting the gauge choice
\begin{align}
  \label{eq:dilfix}
  \phi = 1 \,,
\end{align}
one ends up with the following Lagrangian:
\begin{align}
\label{eq:elCarrgravL}
  \mathcal{L} &= - \frac{w^2}{2} e b^{A} b_{A} + e \pi^{AB}{}_{a} e_{AB}{}^{a} = -\frac{(D-2)^2}{8 (D-p-1)^2} e e^{A}{}_{a}{}^{a} e_{A b}{}^{b} + e \pi^{AB}{}_{a} e_{AB}{}^{a} \,.
\end{align}
Note that for $p=0$, the boost variation \eqref{eq:deltaL0elconf} is zero and consequently there is no need to introduce an extra Lagrange multiplier to make \eqref{eq:L0elconf} invariant.  In that case, the above Lagrangian \eqref{eq:elCarrgravL} lacks the last term and coincides with the electric-type Carroll gravity Lagrangian that was recovered from the conformal approach in \cite{Bergshoeff:2024ilz}. The Lagrangian \eqref{eq:elCarrgravL} can thus be viewed as a generalization of this electric Carroll gravity theory to arbitrary values of $p$.

\subsection{Magnetic $p$-brane Carroll Gravity}

Let us now see whether a $p$-brane Carroll gravity theory can be obtained in a similar manner by starting from the magnetic Lagrangian \eqref{eq:Lflatmagnscal}. Replacing the ordinary derivatives in \eqref{eq:Lflatmagnscal} by the covariant ones, given in \eqref{eq:covderphi}, we obtain the following Ansatz Lagrangian
\begin{align}
  \label{eq:L0magnconf}
  \mathcal{L}_{0, \text{magn. conf.}} &= e \pi^{A} D_{A} \phi - \frac12 e D^{a} \phi D_{a} \phi \,.
\end{align}
This Ansatz is not invariant under homogeneous $p$-brane conformal Carroll transformations but rather transforms as
\begin{align}
  \label{eq:deltaL0magnconf}
  \delta \mathcal{L}_{0, \text{magn. conf.}} &= e D_{A} \phi \left( \delta \pi^{A} - \frac{D}{2} \lambda_D \pi^{A} + \lambda^{A}{}_{B} \pi^{B} - \lambda^{A}{}_{a} D^{a} \phi\right) + \frac{w}{D-p-1} e \lambda^{B}{}_{a} \pi^{A} e_{AB}{}^{a} \phi \nonumber \\ & \qquad + w e \lambda_{K}{}^{a} D_{a}\left(\phi^2\right) \,.
\end{align}
Here, we have taken the weight $w$ as in \eqref{eq:w}, since the second term of \eqref{eq:L0magnconf} is then invariant under dilatations. The last term of \eqref{eq:deltaL0magnconf} shows in particular that $\mathcal{L}_{0, \text{magn. conf.}}$ is not invariant under transversal special conformal transformations. To see how this can be cured, we first partially integrate this term:
\begin{align}
  \label{eq:deltaL0magnconf2}
  \delta \mathcal{L}_{0, \text{magn. conf.}} &= e D_{A} \phi \left( \delta \pi^{A} - \frac{D}{2} \lambda_D \pi^{A} + \lambda^{A}{}_{B} \pi^{B} - \lambda^{A}{}_{a} D^{a} \phi\right) + \frac{w}{D-p-1} e \lambda^{B}{}_{a} \pi^{A} e_{AB}{}^{a} \phi \nonumber \\ & \qquad - w e \omega_{A}{}^{A}{}_{a} \lambda_K{}^{a} \phi^2 - w e \left(D_{a} \lambda_K{}^{a}\right) \phi^2 \,,
\end{align}
where we have defined
\begin{align}
  D_{\mu} \lambda_K{}^{a} \equiv \partial_{\mu} \lambda_{K}{}^{a} + \omega_{\mu}{}^{a}{}_{b} \lambda_{K}{}^{b} - \lambda_{K}{}^{a} b_{\mu} \,.
\end{align}
The last two terms in \eqref{eq:deltaL0magnconf2} can be cancelled by adding the term $w e \left(f_{A}{}^{A} + g_{a}{}^{a}\right) \phi^2$ to $\mathcal{L}_{0, \text{magn. conf.}}$. We thus correct the Ansatz \eqref{eq:L0magnconf} to
\begin{align}
  \label{eq:L1magnconf}
  \mathcal{L}_{1, \text{magn. conf.}} &= e \pi^{A} D_{A} \phi - \frac12 e D^{a} \phi D_{a} \phi + w e \left(f_{A}{}^{A} + g_{a}{}^{a}\right) \phi^2 \,.
\end{align}
This transforms as follows under homogeneous $p$-brane conformal Carroll transformations:
\begin{align}
  \delta \mathcal{L}_{1, \text{magn. conf.}} &= e D_{A} \phi \left( \delta \pi^{A} - \frac{D}{2} \lambda_D \pi^{A} + \lambda^{A}{}_{B} \pi^{B} - \lambda^{A}{}_{a} D^{a} \phi\right) + \frac{w}{D-p-1} e \lambda^{B}{}_{a} \pi^{A} e_{AB}{}^{a} \phi \nonumber \\ & \qquad + w e \left(\Delta f_{A}{}^{A} + \Delta g_{a}{}^{a}\right) \phi^2 + w e \left(D_{A} \lambda_K{}^{A} \right) \phi^2 \,.
\end{align}
Upon partial integration of the last term, this variation can be rewritten as
\begin{align}
  \label{eq:deltaL1magnconf}
  \delta \mathcal{L}_{1, \text{magn. conf.}} &= e D_{A} \phi \left( \delta \pi^{A} - \frac{D}{2} \lambda_D \pi^{A} + \lambda^{A}{}_{B} \pi^{B} - \lambda^{A}{}_{a} D^{a} \phi - 2 w \lambda_K{}^{A} \phi\right) \nonumber \\ & \qquad + \frac{w}{D-p-1} e \lambda^{B}{}_{a} \pi^{A} e_{AB}{}^{a} \phi + w e \left(\Delta f_{A}{}^{A} + \Delta g_{a}{}^{a}\right) \phi^2 \,.
\end{align}
The first line can be set to zero by choosing the transformation rule of $\pi^{A}$ as follows:
\begin{align}
  \label{eq:pitrafo}
  \delta \pi^{A} &= -\lambda^{A}{}_{B} \pi^{B} + \lambda^{A}{}_{a} D^{a} \phi + \frac{D}{2} \lambda_D \pi^{A} + 2 w \lambda_K{}^{A} \phi \,.
\end{align}
For $p=0$, the second line vanishes as well, since in that case the first term is absent and $\Delta f_{A}{}^{A} + \Delta g_{a}{}^{a} = 0$, as mentioned below \eqref{eq:Deltafgtraces}. As was shown in \cite{Bergshoeff:2024ilz}, the Lagrangian \eqref{eq:L1magnconf} is then gauge equivalent to that of $0$-brane magnetic Carroll gravity in the second-order formulation. For non-zero values of $p$, the second line of \eqref{eq:pitrafo} no longer vanishes and $\mathcal{L}_{1, \text{magn. conf.}}$ is no longer boost invariant. Since it is not clear how boost invariance can be restored (either with the field content at hand or by introducing extra auxiliary fields), the Lagrangian \eqref{eq:L1magnconf} can not be used in a straightforward manner to construct $p$-brane magnetic Carroll gravity for $p\neq 0$. Furthermore, from \eqref{eq:depfgsols} one sees that the combination $f_{A}{}^{A} + g_{a}{}^{a}$, which would remain after gauge fixing dilatations and special conformal transformations, does not reproduce the $R_{AB}(J^{AB})$ curvature term of the $p$-brane magnetic Carroll gravity Lagrangian for $p>0$. For these reasons, we will no longer consider the $p$-brane conformal Carroll algebra for this purpose. Instead, in the next section, we will show how for non-zero values of $p$, magnetic Carroll gravity can be obtained from the conformal approach by starting from a different conformal algebra.

\section{A Conformal approach to  $p$-brane Carroll gravity for $p>1$ }\label{sec:pconfargrav}

In this section, we will revisit the conformal approach to the construction of the Lagrangian of magnetic $p$-brane Carroll gravity in the second-order formulation for $p \neq 0,1$. Instead of gauging and implementing the homogeneous $p$-brane conformal Carroll symmetries, we will start from a different algebra of homogeneous conformal Carroll transformations. This algebra contains longitudinal Lorentz transformations $J_{AB}$, a longitudinal dilatation $D$, longitudinal special conformal transformations $K_{A}$, transversal rotations $J_{ab}$ and a second dilatation $\tilde{D}$ that acts anisotropically on both longitudinal and transversal directions. 
We introduce gauge fields $\omega_{\mu}{}^{AB}$, $b_{\mu}$, $f_{\mu}{}^{A}$, $\omega_{\mu}{}^{ab}$ and $c_{\mu}$ that are associated to these respective generators, as well as longitudinal and transversal Vielbeine $\tau_{\mu}{}^{A}$ and $e_{\mu}{}^{a}$. We then posit that these fields have the following transformation rules:
\begin{align}
    \label{eq:trafosnewalg}
    \delta \tau_{\mu}{}^{A} &= - \lambda^{A}{}_{B} \tau_{\mu}{}^{B} - \lambda_D \tau_{\mu}{}^{A} + \sigma \tau_{\mu}{}^{A} \,, \nonumber \\
    \delta e_{\mu}{}^{a} &= - \lambda^{a}{}_{b} e_{\mu}{}^{b} - \frac{(p-1)}{(D-p-1)} \sigma e_{\mu}{}^{a} \,, \nonumber \\
    \delta \omega_{\mu}{}^{AB} &= \partial_{\mu} \lambda^{AB} - 2 \lambda^{[A|C|} \omega_{\mu C}{}^{B]} - 4 \lambda_K{}^{[A} \tau_{\mu}{}^{B]} \,, \nonumber \\
    \delta \omega_{\mu}{}^{ab} &= \partial_{\mu} \lambda^{ab} - 2 \lambda^{[a|c|} \omega_{\mu c}{}^{b]} \,, \nonumber \\
    \delta f_{\mu}{}^{A} &= \partial_{\mu} \lambda_K{}^{A} + \omega_{\mu}{}^{AB} \lambda_{K B} - \lambda^{A}{}_{B} f_{\mu}{}^{B} - \lambda_K{}^{A} b_{\mu} + \lambda_D f_{\mu}{}^{A} - \sigma f_{\mu}{}^{A} + \lambda_K{}^{A} c_{\mu} \,, \nonumber \\
    \delta b_{\mu} &= \partial_{\mu} \lambda_D + 2 \lambda_K{}^{A} \tau_{\mu A} \,, \nonumber \\
    \delta c_{\mu} &= \partial_{\mu} \sigma \,,
\end{align}
where $\lambda^{AB}$, $\lambda_D$, $\lambda_{K}{}^{A}$, $\lambda^{ab}$ and $\sigma$ are the parameters of the $J_{AB}$, $D$, $K_{A}$, $J_{ab}$ and $\tilde{D}$ transformations. In what follows, we will need the following covariant field strengths:
\begin{subequations}
\label{eq:curvsnewalg}
\begin{align}
    R_{\mu\nu}(P^{A}) &= 2 \partial_{[\mu} \tau_{\nu]}{}^{A} + 2 \omega_{[\mu}{}^{AB} \tau_{\nu] B} + 2 b_{[\mu} \tau_{\nu]}{}^{A} - 2 c_{[\mu} \tau_{\nu]}{}^{A} \,, \label{eq:RPAnewalg} \\
     R_{\mu\nu}(P^{a}) &= 2 \partial_{[\mu} e_{\nu]}{}^{a} + 2 \omega_{[\mu}{}^{ab} e_{\nu] b} + 2 \frac{(p-1)}{(D-p-1)} c_{[\mu} e_{\nu]}{}^{a} \,, \label{eq:RPanewalg} \\
     \mathcal{R}_{\mu\nu}(J^{AB}) &= 2 \partial_{[\mu} \omega_{\nu]}{}^{AB} + 2 \omega_{[\mu}{}^{[A|C|} \omega_{\nu] C}{}^{B]} + 8 f_{[\mu}{}^{[A} \tau_{\nu]}{}^{B]} \,. \label{eq:RMnewalg} 
\end{align}
\end{subequations}

We now turn to explaining how this conformal algebra can be used to obtain the Lagrangian for $p$-brane Carroll gravity in the second-order formulation. In what follows, we will assume that $p$ is not equal to 0 or 1. As remarked in section \ref{sec:pgravreview}, the case $p=0$ is special in the sense that its Lagrangian has a different form than that of the $p\neq 0$ cases. It has been treated in detail in \cite{Bergshoeff:2024ilz}. Even though its Lagrangian resembles that of all other $p \neq 0$ cases, the $p=1$ case is also special, since then the group of longitudinal Lorentz transformations is abelian. As remarked in \cite{Bergshoeff:2023rkk}, this implies that there is no second-order formulation and all spin connection components appear as independent Lagrange multipliers in the Lagrangian. Since the conformal approach we have in mind leads to second-order actions, we will not treat this case.\,
We will see that some of the formulas below are not well-defined when $p=0$ or $p=1$, confirming the fact that these are particular cases for which the procedure of this section does not work.

We can use the curvatures \eqref{eq:curvsnewalg} to impose the following three conventional constraints:
\begin{align}
    \label{eq:convconstrnewalg}
    R_{Aa}(P^{a}) = 0 \,, \qquad \qquad R_{BC}(P^{A}) = 0 \,, \qquad \qquad \mathcal{R}_{AB}(J^{AB}) = 0 \,.
\end{align}
The first of these can be solved for the longitudinal components of the $\tilde{D}$ gauge field $c_{\mu}$:
\begin{align}
    \label{eq:solcnewalg}
    c_{A} &= - \frac{1}{(p-1)} e_{Aa}{}^{a} \,.
\end{align}
The second constraint of \eqref{eq:convconstrnewalg} can be used to solve for the longitudinal components of $\omega_{\mu}{}^{AB}$:
\begin{align}
    \label{eq:solomnewalg}
    \omega_{C,AB} &= \tau^\prime_{C[A,B]} - \frac12 \tau^{\prime}_{AB,C} + \frac{2}{(p-1)} e_{a[A}{}^{a} \eta_{B]C} \,,
\end{align}
where we have used \eqref{eq:solcnewalg} and we have introduced the notation
\begin{align}
    \label{eq:deftauprime}
    \tau^{\prime}_{\mu\nu}{}^{A} \equiv 2 \partial_{[\mu} \tau_{\nu]}{}^{A} + 2 b_{[\mu} \tau_{\nu]}{}^{A} \,.
\end{align}
Note that we only solve for the longitudinal components $\omega_{C}{}^{AB}$ of $\omega_{\mu}{}^{AB}$ via conventional constraints. We do not impose a conventional constraint that allows to also solve for the transversal components $\omega_{a}{}^{AB}$. The latter remain as independent spin connection components. Finally, the last conventional constraint of \eqref{eq:convconstrnewalg} gives the following solution for the trace of the special conformal gauge field $f_{\mu}{}^{A}$:
\begin{align}
    \label{eq:fAAnewalg}
    f_{A}{}^{A} &= - \frac{1}{4 p} R_{AB}(J^{AB}) \,, \qquad \ \ \ \text{where } \ R_{\mu\nu}(J^{AB}) \equiv 2 \partial_{[\mu} \omega_{\nu]}{}^{AB} + 2 \omega_{[\mu}{}^{[A|C|} \omega_{\nu] C}{}^{B]} \,.
\end{align}

We now consider a scalar field $\phi$ that only transforms under the longitudinal dilatation $D$ with a scaling weight $w$:
\begin{align}
    \delta \phi &= w \lambda_D \phi \,,
\end{align}
and we define its longitudinal covariant derivative $D_{A} \phi$ and longitudinal conformal Laplacian $D^{A} D_{A} \phi$ as
\begin{align}
    D_{A} \phi &= \tau_{A}{}^{\mu} \left( \partial_{\mu} \phi - w b_{\mu} \phi \right) \,, \nonumber \\
    D^{A} D_{A} \phi &= \tau^{A \mu} \left(\partial_{\mu} D_{A} \phi + \omega_{\mu A}{}^{B} D_{B} \phi - (w + 1) b_{\mu} D_{A} \phi + c_{\mu} D_{A} \phi + 2 w f_{\mu A} \phi \right) \,.
\end{align}
These transform as follows:
\begin{align}
    \label{eq:deltaDphilaplphi}
    \delta D_{A} \phi &= - \lambda_{A}{}^{B} D_{B} \phi + (w + 1) \lambda_D D_{A} \phi - \sigma D_{A} \phi - 2 w \lambda_{K A} \phi \,, \nonumber \\
    \delta D^{A} D_{A} \phi &= (w + 2) \lambda_D D^{A} D_{A} \phi - 2 \sigma D^{A} D_{A} \phi - 2 \left(2 w - (p - 1) \right) \lambda_K{}^{A} D_{A} \phi \,.
\end{align}
The following Lagrangian
\begin{align}
    \label{eq:confLnewalg}
    \mathcal{L} = e \phi D^{A} D_{A} \phi \,,
\end{align}
is then invariant under all transformations $J_{AB}$, $J_{ab}$, $D$, $\tilde{D}$ and $K_{A}$, provided we choose $w$ as
\begin{align}
    w = \frac{(p-1)}{2} \,.
\end{align}
Indeed, invariance under $J_{AB}$ and $J_{ab}$ is trivial. Invariance of \eqref{eq:confLnewalg} under $K_{A}$ follows from the choice of $w$, which makes the $K_{A}$ transformation of $D^{A} D_{A} \phi$ vanish (see \eqref{eq:deltaDphilaplphi}). This choice of $w$ also makes \eqref{eq:confLnewalg} invariant under the longitudinal dilatation $D$. Finally, invariance under the anisotropic dilatation $\tilde{D}$ follows from the fact that $e$ and $D^{A} D_{A} \phi$ scale with weights 2 and $-2$ respectively, while $\phi$ itself is invariant under $\tilde{D}$. 

The longitudinal dilatation $D$ and special conformal transformations $K_{A}$ can be fixed by adopting the following gauge choices:
\begin{align}
    D \ \text{gauge } : \ \phi = 1 \,, \qquad \qquad \qquad K_{A} \ \text{gauge } : \ b_{A} = 0 \,.
\end{align}
The Lagrangian \eqref{eq:confLnewalg} then reduces to
\begin{align}
    \label{eq:Lgfnewalg}
    \mathcal{L}^\prime &= (p-1) e f_{A}{}^{A} = - \frac{(p-1)}{4 p} e R_{AB}(J^{AB}) \,.
\end{align}
Note that the longitudinal components of the spin connection $\omega_{\mu}{}^{AB}$ that appear in $R_{AB}(J^{AB})$ are now given by (see \eqref{eq:solomnewalg} with $b_{A} = 0$):
\begin{align}
    \omega_{C,AB} &= \tau_{C[A,B]} - \frac12 \tau_{AB,C} + \frac{2}{(p-1)} e_{a[A}{}^{a} \eta_{B]C} \,.
\end{align}
The transversal components $\omega_{a}{}^{AB}$ remain independent. Up to an overall prefactor, the Lagrangian \eqref{eq:Lgfnewalg} then coincides with the Lagrangian of $p$-brane magnetic Carroll gravity of \cite{Bergshoeff:2023rkk} in the second-order formulation.

This concludes our discussion on the conformal approach for Carroll gravity theories. As mentioned in the introduction, there is a similar discussion for dual $p$-brane Galilei gravities. For convenience of the reader, we will treat this case in the Appendices.
\section{Conclusions and Outlook} \label{sec:outlook}
In this work we showed how the conformal technique can be applied to any $p$-brane foliated Carroll symmetry and, by brane-duality, to any $p$-brane foliated Galilei symmetry. In the magnetic case, we found the general rule that, if the Carroll gravity theory does not contain a boost spin connection, the required conformal extension of the Carroll algebra is a conformal extension in the longitudinal directions only supplemented with an additional an-isotropic dilatation to explain for the emergent dilatations we find in the final result. 
This algebra is not the $p$-brane limit of the relativistic conformal algebra. The Carroll boost symmetry  arises as an emerging symmetry after the conformal coupling. We also learned that since some of the spin connection fields remain independent in a second-order  formulation, in order to have an action formulation, we should not impose the maximum set of conventional constraints. These independent spin connection fields occur as Lagrange multipliers in the action leading to constraints on the geometry. We also showed that the $p$-brane conformal Carroll algebra can be used to construct examples of  electric Carroll gravity theories as in \eqref{eq:elCarrgravL}. In contrast, the new conformal algebra cannot be used to construct boost-covariant intrinsic torsion tensors, thereby making the construction of new electric Carroll gravity theories based on this new conformal algebra not obvious. 

On the other hand, if the Carroll gravity theory does contain a Carroll boost, the required conformal extension is the $p$-brane limit of the relativistic conformal algebra which we called the $p$-brane conformal Carroll algebra. Again, some components of the Carroll boost spin connection are independent leading to constraints on the geometry but in this case they could anyhow not be solved for by imposing a maximum set of conventional constraints.  The corresponding statements for the Galilei case can be found in the Appendices. In that case, the domain wall  with $p=D-2$ is exceptional.
The string Carroll  with $p=1$ and the Galilei defect brane   with $p=D-3$ are exceptional in two respects. First of all,  the corresponding gravity theories are already invariant under local dilatations and therefore there is nothing to compensate for. This is analogous to four-dimensional conformal gravity  in the relativistic case. Secondly, for these special cases the action does not contain any term quadratic in the spin connections and therefore all  spin connection fields are independent. Consequently,  these special cases do not have  a second-order formulation. 

The different gravitational actions of the magnetic type we constructed in this work all follow  from the coupling of a scalar to conformal gravity. In total we discussed $4=2 \times 2 $ different actions. The first factor of $2$ corresponds to the fact that for every Carroll action, by brane duality, we have a dual Galilei action. We discussed these dual Galilei actions in two separate Appendices. The second factor of $2$ is related to the fact that there is always a special case (particle for Carroll and domain wall for Galilei) and a generic $p$-brane case with different conformal algebras. 

In the magnetic case, the coupling to conformal gravity follows the following pattern. We first consider the generic $p$-brane case. As a first step, we replace the ordinary derivative of the scalar by a dilatation-covariant derivative. 
We always find that the longitudinal dilatation gauge field components that appear, are independent and transform under a special conformal transformation. The dilatation gauge field is then not a connection and this makes the construction of a covariant d'Alembertian non trivial. As in the relativistic case, one needs to add to the action a term that is linear in the dependent components of the (trace of the) special conformal transformations gauge field. In all cases this dependent expression is proportional to a  curvature tensor and this leads to a `magnetic' gravitational action involving a curvature tensor of longitudinal or transversal rotations.  In the special magnetic cases, the procedure follows a similar pattern except that  now the dependent expressions of {\sl two} (longitudinal and transversal) special conformal gauge fields are involved. Via a slightly more involved calculation,  they lead to a boost-invariant combination of the two curvature tensors that make up a magnetic gravity action for the special case. To obtain this boost-invariant combination requires a fine-tuning that only works for the special case but is not valid in the generic case. This  motivated us to use a different conformal algebra for the generic case.

It is interesting to compare the pattern we found in this work  with the Aristotelian case where there are no boosts at all. We would then only have a generic case where  the relevant conformal extension is an extension separately in both the longitudinal and transversal directions with, in general, two different dilatations to be used in the compensating mechanism supplemented with two more an-isotropic dilatations to explain for the  two emergent dilatations that arise in the separate Carroll and Galilei parts that make up  the Aristotelian gravity action. This would lead to a slightly different construction of magnetic Aristotelian gravity as compared with  \cite{Bergshoeff:2025qtt}. The difference relies in the fact that in  \cite{Bergshoeff:2025qtt} we did not require the existence of a first-order formulation neither did the final magnetic Aristotelian gravity Lagrangian contain Lagrange multiplier fields given by independent spin-connection fields.
We note that both  the Aristotelian algebra as well as its conformal extension cannot be obtained by taking any $p$-brane limit of the corresponding relativistic algebras.

 In section 3.1 we also  found that for a flat spacetime, the $p$-brane conformal Carroll isometries formed an infinite-dimensional extension of the $p$-brane conformal Carroll transformations involving supertranslations and superrotations. It would be interesting to see in which sense these isometries are related to foliated BMS symmetries, i.e.~the asymptotic isometries of an asymptotically flat spacetime as probed by a $p$-brane instead of a particle. This should lead to a rather different analysis than the particle case. 

Finally, we remark that the original goal of the conformal technique was to have a systematic way to construct Carroll matter couplings. A first exercise could be to construct a Carroll version of NS-NS gravity, and possibly its supersymmetric extension making use of  Carroll fermions \cite{Bagchi:2022eui, Bergshoeff:2023vfd, Bergshoeff:2024ytq, Grumiller:2025rtm}. This method should be compared with the alternative technique of taking a $p$-brane Carroll limit of a relativistic matter coupling. We hope to come back to such Carroll matter couplings in a future work.

\section*{Acknowledgements}

E.R. acknowledges financial support from ANID through Fondecyt grant No. 1250642 and 1231133. This work was
supported by the Research project Code DIREG 04/2025 (P.C.) of the Universidad Católica de la Santísima
Concepción (UCSC), Chile. P.C. and E.R. would like to thank to the Direcci\'{o}n de Investigaci\'{o}n and Vicerector\'{\i}a de Investigaci\'{o}n of the Universidad Cat\'{o}lica de la Sant\'{\i}sima Concepci\'{o}n, Chile, for their constant support. This work is supported by the Croatian Science Foundation project IP-2022-10-5980 “Non-relativistic supergravity and applications”  and by the European Union–NextGenerationEU. The views and opinions expressed are those of the authors only and do not necessarily reflect the official views of the European Union or the European Commission. Neither the European Union nor the European Commission can be held responsible for them. 

\appendix
\section{$p$-brane  Galilei gravity for $p<D-2$}\label{appendixa}
In this section, we present a conformal approach to 
$p$-brane Galilei gravity for $p \ne D-2$. The central idea is to first gauge a related conformal algebra that does not include boosts. This approach is motivated by the observation that the conventional conformal construction works  well when only rotations are present, as in the cases of general relativity and Aristotelian gravity \cite{Bergshoeff:2025qtt}. In this Appendix we consider the conformal approach to the general $p$-brane Galilei gravity case except for the case that $p=D-2$, as it requires a different conformal extension that we will consider in Appendix B. In that case, our starting point is the $p$-brane Galilean conformal algebra  that can be obtained as a domain wall limit of the relativistic conformal algebra.

Following the reasoning outlined above, we propose that in the Galilean case the appropriate starting point should be the homogeneous conformal algebra
including transversal rotations $J_{ab}$, transversal special conformal transformations $K_a$, a transversal dilatation $D$, longitudinal Lorentz transformations $M_{AB}$ and a
 second dilatation $\tilde{D}$ that acts anisotropically on both longitudinal and transversal directions.
Along these generators, we introduce the associated gauge fields $\omega_{\mu}{}^{ab}$, $f_\mu{}^{a}$, $b_{\mu}$, $\omega_{\mu}{}^{AB}$ and $c_{\mu}$, as well as transversal and longitudinal Vielbeine $e_\mu{}^{a}$ and $\tau_\mu{}^{A}$. We then assume that these fields transform as follows
\begin{align}
  \delta \tau_{\mu}{}^{A} &=- \lambda^{A}{}_{B} \tau_{\mu}{}^{B} - \frac{(D-p-3)}{(p+1)} \sigma \tau_{\mu}{}^{A}  \,, \notag  \\
  \delta e_{\mu}{}^{a} &=  -\lambda^{a}{}_{b} e_{\mu}{}^{b} - \lambda_D e_{\mu}{}^{a} +\sigma e_{\mu}{}^{a}\,,\notag \\
   \delta \omega_{\mu}{}^{AB} &= \partial_{\mu} \lambda^{AB} - 2 \lambda^{[A|C|} \omega_{\mu C}{}^{B]}  \,, \notag \\
  \delta \omega_{\mu}{}^{ab} &= \partial_{\mu} \lambda^{ab} - 2 \lambda^{[a|c|} \omega_{\mu c}{}^{b]} - 4 \lambda_{K}{}^{[a} e_{\mu}{}^{b]} \,, \notag  \\
  \delta f_{\mu}{}^{a} &= \partial_{\mu} \lambda_{K}{}^{a} - \lambda_{K b}  \omega_{\mu}{}^{ba}- \lambda_{K}{}^{a} b_{\mu} - \lambda^{a}{}_{b} f_{\mu}{}^{b}  +\lambda_D f_{\mu}{}^{a}- \sigma f_{\mu}{}^{a} + \lambda_K{}^{a} c_{\mu} \,, \notag  \\
  \delta b_{\mu} &= \partial_{\mu} \lambda_D + 2 \lambda_{K}{}^{a} e_{\mu a} \,,\notag\\
 \delta c_{\mu} &=\partial_\mu \sigma \,, \label{eq:haconfrules0}
\end{align}
 with respective gauge parameters $\lambda^{ab},\lambda_K{}^{a},\lambda_D,\lambda^{AB}$ and $\sigma$. 
An appealing feature of these transformation rules is that they give rise to a spatial rotation curvature containing a special conformal gauge field. The absence of such a term was previously identified as an obstacle to gauging Galilei gravity \cite{Bergshoeff:2024ilz}.  In what follows, we will make use of the following covariant field strengths:
\begin{subequations}
\label{eq:carrconfcurvs}
\begin{align}   
  R_{\mu\nu}(P^A) &= 2\, \partial_{[\mu} \tau_{\nu]}{}^{A} + 2\, \omega_{[\mu}{}^{AB} \, \tau_{\nu] B}  +2\frac{(D-p-3)}{(p+1)}c_{[\mu}\tau_{\nu]}{}^{A}\,,\label{RPAGalp}  \\
  R_{\mu\nu}(P^{a}) &= 2\, \partial_{[\mu} e_{\nu]}{}^{a} + 2\, \omega_{[\mu}{}^{ab} \, e_{\nu] b} + 2\, b_{[\mu} \, e_{\nu]}{}^{a} -2\, c_{[\mu} \, e_{\nu]}{}^{a}\,,  \label{RPaGalp}\\
  R_{\mu\nu}(J^{ab}) &= 2 \, \partial_{[\mu} \omega_{\nu]}{}^{ab} + 2 \, \omega_{[\mu|}{}^{[a|}{}_{c} \, \omega_{|\nu]}{}^{c|b]} + 8 \, f_{[\mu}{}^{[a} \, e_{\nu]}{}^{B]}\,.\label{RJabGalp}
\end{align}
\end{subequations}
Up to this point, all gauge fields $\tau_{\mu}{}^{A},\omega_{\mu}{}^{AB},e_{\mu}{}^{a},\omega_{\mu}{}^{ab},f_{\mu}{}^{a},b_\mu$ and $c_\mu$ have been considered as independent. However, one can impose conditions that render a subset of these fields dependent. To realize this, we impose, as in the Carroll case,  suitable curvature constraints that allow certain fields to be expressed in terms of the remaining ones, thereby reducing the number of independent fields. 
Inspection of the curvatures listed in Eq.\eqref{eq:carrconfcurvs} leads us to impose the following three conventional constraints:
\begin{align}
    \label{eq:convconstrgalp}
    R_{aA}(P^{A}) = 0 \,, \qquad \qquad R_{bc}(P^{a}) = 0 \,, \qquad \qquad {R}_{ab}(J^{ab}) = 0 \,.
\end{align}
Considering the first constraint in \eqref{eq:convconstrgalp}, we find the transversal component of the gauge field $c_\mu$,
\begin{align}
    \label{eq:solcagalp}
    c_{a} &= - \frac{1}{(D-p-3)} \tau_{aA}{}^{A} \,.
\end{align}
From the second constraint we find the transversal component of $\omega_\mu{}^{ab}$,
\begin{equation}
\omega_{c,ab}=e^{\prime}_{c[a,b]}-\frac{1}{2}e^{\prime}_{ab,c}+\frac{2}{(D-p-3)}\tau_{A[a}{}^{A}\delta_{b]c}\,,\label{omegacab}
\end{equation}
where we have used \eqref{eq:solcagalp} and we have defined 
\begin{equation}
    e^{\prime}_{\mu\nu}{}^{a}\equiv 2 \partial_{[\mu}e_{\nu]}{}^{a}+2b_{[\mu}e_{\nu]}{}^{a}\,.
\end{equation}
We stress that conventional constraints are
used only to solve for the transversal components $\omega_{c}{}^{ab}$ of $\omega_{\mu}{}^{ab}$. No conventional constraint is imposed to determine the longitudinal components $\omega_{C}{}^{ab}$ which therefore remain as independent components. 
 The conventional constraint $R_{a b}(J^{ab})=0$ can be used to determine the trace of the transversal special conformal gauge field $f_\mu{}^a$, which is found to be:
\begin{equation}\label{solution}
f_a{}^a = -\frac{1}{4 (D-p-2)} R'_{ab}{}(J^{ab})\,.
\end{equation}
Here, $R^\prime_{\mu\nu}(J^{ab})$ is given by the expression 
\begin{equation}\label{R(J)}
    R^\prime_{\mu\nu}(J^{ab})= 2 \, \partial_{[\mu} \omega_{\nu]}{}^{ab} + 2 \, \omega_{[\mu|}{}^{[a|}{}_{c} \, \omega_{|\nu]}{}^{c|b]}\,.
\end{equation}

\subsection{A conformal approach to $p$-brane Galilei gravity for $p<D-2$}
We now consider a scalar field $\phi$ whose transformation rule  under longitudinal dilatation $D$ and scaling weight $w$ is given by:
\begin{equation}
    \delta \phi=w \lambda_D\phi\,.\label{delta_phigalp}
\end{equation}
We also define its transversal covariant derivative and transversal conformal Laplacian $D^aD_a\phi$ as follows: \begin{align}
 D_a\phi&= e_a{}^{\mu}\left(\partial_\mu\phi-w b_{\mu}\phi\right)\,, \notag \\
 D^aD_a\phi&=e^{\mu a}\left(\partial_\mu D_a\phi-(w+1)b_\mu D_a\phi+c_\mu D_a\phi+\omega_{\mu a}{}^{b}D_b \phi+2w f_{\mu a}\phi\right)\,,\label{DaphiDaDaphi}
\end{align}
that transform as:
\begin{align}
    \label{eq:deltaDphilaplphigalp}
    \delta D_{a} \phi &= - \lambda_{a}{}^{b} D_{b} \phi + (w + 1) \lambda_D D_{a} \phi - \sigma D_{a} \phi - 2 w \lambda_{K a} \phi \,, \nonumber \\
    \delta D^{a} D_{a} \phi &= (w + 2) \lambda_D D^{a} D_{a} \phi - 2 \sigma D^{a} D_{a} \phi -\left[2(D-p-1)-4w-4\right]\lambda_K{}^{a}D_a\phi \,.
\end{align}

With the covariant derivative and the transversal conformal Laplacian defined in \eqref{DaphiDaDaphi}, we can now construct a Lagrangian for the scalar field $\phi$, that is invariant under the $J_{AB}$, $J_{ab}$, $D$, $\tilde{D}$ and $K_{a}$ transformations. We then find the following invariant Lagrangian:
\begin{align}
    \label{eq:Lconfmagnmagn}
    \mathcal L_{\rm conformal} = \frac{1}{2}  e  \phi D^a D_a \phi  \,,
\end{align}
provided we choose $w$ as
\begin{equation}
w =\frac{1}{2}(D-p-3)\,.\label{ww}
\end{equation}
In fact, this choice of $w$ makes the Lagrangian invariant under the $K_a$ transformations as well as the transversal dilatation $D$, since the determinant $e$ has weight $-(D-p-1)$ . By construction, invariance under $J_{AB}$ and $J_{ab}$ is ensured. Besides, invariance under the anisotropic dilation $\tilde{D}$ is also satisfied since $\phi$ is invariant and $e$ scales with weight 2.

The transversal dilatation $D$ and transversal special conformal transformations $K_{a}$ can be fixed by the choice:
\begin{align}
    D \ \text{gauge } : \ \phi = 1 \,, \qquad \qquad \qquad K_{a} \ \text{gauge } : \ b_{a} = 0 \,.
\end{align}
Using the solution \eqref{solution} for $f_a{}^a$ then leads to the following Lagrangian
\begin{align} \label{eq:fixedmagnLag}
 \mathcal{L}^{\prime} =\frac{1}{2}e(D-p-3)f_a{}^{a}= -\frac{1}{8}\frac{(D-p-3)}{(D-p-2)} \,  e\,  R_{ab}{}^{ab}(J)  \,.
\end{align}
The transversal component of the spin connection $\omega_{\mu}{}^{ab}$ appearing in $R_{ab}(J^{ab})$ are now given by: \begin{align}
\omega_{c,ab} &= e_{c[a,b]} - \frac{1}{2} e_{ab,c} + \frac{2}{(D-p-3)} \tau_{A[a}{}^{A} \delta_{b]c} \,,
\end{align}
that is \eqref{omegacab} with $b_a=0$. The components $\omega_{C}{}^{ab}$ along the longitudinal directions remain independent. Up to an overall prefactor, the Lagrangian \eqref{eq:fixedmagnLag} coincides with that of \cite{Bergshoeff:2023rkk}, describing the first-order $p$-brane Galilei Lagrangian.

\section{  Domain wall Galilei  Gravity}
A special example of the duality between  Galilei geometry and Carroll geometry (from a brane perspective) is that domain wall Galilean gravity is dual to particle Carroll gravity. The conformal mechanism was already applied to the particle Carroll gravity in \cite{Bergshoeff:2024ilz}. Although this case is dual to the conformal domain wall Galilei  gravity construction, here we explicitly illustrate how the conformal mechanism is implemented in domain wall $p$-brane Galilean gravity.

\subsection{The conformal domain wall Galilean algebra} \label{ssecB:pconfcarralg}
The conformal domain wall Galilean algebra denoted here as $\mathfrak{cgal}_{\text{dw}}$ is spanned by the following generators:
\begin{equation}
    \mathfrak{cgal}_{\text{dw}}=\text{Span}\lbrace P_A, P, J_{AB},G_A,K,K_A,D\rbrace\,.
\end{equation}
Here, the longitudinal index $A,B,\cdots=0,1\cdots,D-2$, while there is only one transversal direction $a=z$, and hence $J_{ab}=0$, $G_A:=J_{zA}$, $P:=P_{z}$, $K:=K_{z}$. $P_A$ and $P$ will be referred to as longitudinal and transversal translation generators, respectively. Moreover, $J_{AB}$ span the Lie algebra $\mathfrak{so} (1,D-2)$ of Lorentz transformations of the longitudinal directions. On the other hand, $G_A$ will be called domain wall Galilean boosts. Together, the generators $P_{A}$, $P$, $J_{AB}$, and $G_{A}$ form the 
domain wall Galilean algebra.  The additional generators that extend this algebra to $\mathfrak{cgal}_{\text{dw}}$ are the longitudinal special conformal generators $K_A$, the transversal special conformal generator $K$ and the dilatation generator $D$.

The non-zero commutation relations of this algebra are given by:
\begin{alignat}{3} \label{eq:CGA1}
\comm{J_{AB}}{J_{CD}} &= 4\, \eta_{[A[C} J_{D]B]} \,, \qquad & \comm{J_{AB}}{G_{C}} &= 2\, \eta_{C[B} G_{ A]} \qquad & \comm{P_{A}}{J_{BC}} &= 2 \, \eta_{A[B} P_{C]} \,, \nonumber \\[.1truecm]
  \comm{P_{A}}{G_{B}} &= - \, \eta_{AB}P \,, \qquad & \comm{P_{A}}{K_{B}} &= 2\, \left(\eta_{AB}D+J_{AB}\right) \,, \qquad & \comm{K_{A}}{J_{BC}} &= 2\,\eta_{A[B} K_{C]} \,, \nonumber \\[.1truecm]
\comm{K_{A}}{G_{B}} &= - \, \eta_{AB}K \,, \qquad &  \comm{P_{A}}{K} &= -2 \, G_{A} \,, \qquad & \comm{P}{K_{A}} &= 2\,G_{A} \,, \nonumber \\[.1truecm]
 \comm{D}{P_{A}} &= P_{A} \,, \qquad & \comm{D}{P} &= P \,, \qquad & \comm{D}{K_{A}} &=- K_{A} \,, \nonumber \\[.1truecm]
\comm{D}{K} &= -K \,.
\end{alignat}
From these commutators we can see that the generators $\lbrace J_{AB},G_A,K,K_A,D\rbrace$ span a subalgebra which we will denote as $\mathfrak{hcgal}_{\text{dw}}$, and that we will call the subalgebra of homogeneous conformal domain wall  Galilean transformations. 

\subsection{Gauging the conformal domain wall Galilean algebra}
The gauging process starts by introducing the $\mathfrak{cgal}_{\text{dw}}$-valued gauge field
\begin{align}
  \label{Acg}
  A_\mu &= \tau_\mu{}^A P_A + e_\mu P + \frac12 \omega_{\mu}{}^{AB} J_{AB} + \omega_{\mu}{}^{A} G_{A} + f_{\mu}{}^{A} K_A + g_{\mu} K + b_\mu D \,,
\end{align}
and gauge parameter
\begin{align}
  \label{Sigmacg}
  \Sigma &= \zeta^A H_A + \zeta P + \frac12 \lambda^{AB} J_{AB} + \lambda^{A} G_{A} +  \lambda_K{}^A K_A + \lambda_K{} K + \lambda_D D \,.
\end{align}
Then, the transformation rules of the component gauge fields $\tau_{\mu}{}^{A}$, $e_{\mu}{}$, $\omega_{\mu}{}^{AB}$, $\omega_{\mu}{}^{A}$, $f_{\mu}{}^{A}$, $g_{\mu}{}$ and $b_{\mu}$ can be derived from the general gauge transformation rule of $A_{\mu}$:
\begin{align}
  \label{deltaAcg}
  \delta A_\mu &= \partial_\mu \Sigma + \commutator{A_{\mu}}{\Sigma} \,.
\end{align}
Analogously to the Carroll case, we will mostly omit the local longitudinal and transversal translations with parameters  $\zeta^{A}$, $\zeta$ and instead concentrate on the homogeneous conformal domain wall Galilei transformations.  We find from \eqref{deltaAcg} that the gauge fields transform under these transformations as follows,
\begin{align}
  \label{eq:confgaltrafos}
  \delta e_\mu &= - \lambda^{A} \tau_{\mu A} - \lambda_D e_\mu \,, \nonumber \\
  \delta \tau_\mu{}^A &=  - \lambda^{A}{}_{B} \tau_\mu{}^{B} - \lambda_D \tau_\mu{}^A \,, \nonumber \\
  \delta \omega_\mu{}^{AB} &= \partial_\mu \lambda^{AB} - 2 \lambda^{[A|}{}_C \omega_\mu{}^{C|B]} - 4 \lambda_K{}^{[A} \tau_\mu{}^{B]}  \,, \nonumber \\
  \delta \omega_\mu{}^{A} &= \partial_\mu \lambda^{A} + \omega_\mu{}^{A}{}_{B} \lambda^{B} - \lambda^{A}{}_{B} \omega_\mu{}^{B} + 2 \lambda_K{}^{A} e_\mu - 2 \lambda_K \tau_\mu{}^{A}\,, \nonumber \\
  \delta g_\mu &= \partial_\mu \lambda_K - \lambda_K b_\mu - \lambda^{A} f_{\mu A} + \lambda_{K A} \omega_\mu{}^{A} + \lambda_D g_\mu \,, \nonumber \\
  \delta f_\mu{}^A &= \partial_\mu \lambda_K{}^A + \omega_\mu{}^{A}{}_{B} \lambda_K{}^B - \lambda_K{}^{A} b_\mu - \lambda^{A}{}_{B} f_\mu{}^{B} + \lambda_D f_\mu{}^{A} \,, \nonumber \\
  \delta b_\mu &= \partial_\mu \lambda_D + 2 \lambda_K{}^{A} \tau_{\mu A}  \,.
\end{align}
We also define the following curvatures,
\begin{subequations} \label{eq:cgalconfcurvs}
\begin{align}   
  R_{\mu\nu}(P) &= 2\, \partial_{[\mu} e_{\nu]}{} + 2\, \omega_{[\mu}{}^{A} \, \tau_{\nu] A} + 2\, b_{[\mu} \, e_{\nu]} \,,  \\
  R_{\mu\nu}(P^{A}) &= 2\, \partial_{[\mu} \tau_{\nu]}{}^{A} + 2\, \omega_{[\mu}{}^{AB} \, \tau_{\nu] B} + 2\, b_{[\mu} \, \tau_{\nu]}{}^{A} \,,  \\
  R_{\mu\nu}(J^{AB}) &= 2 \, \partial_{[\mu} \omega_{\nu]}{}^{AB} + 2 \, \omega_{[\mu|}{}^{[A|}{}_{C} \, \omega_{|\nu]}{}^{C|B]} + 8 \, f_{[\mu}{}^{[A} \, \tau_{\nu]}{}^{B]} \,, \label{eq:RJAB} \\
  R_{\mu\nu}(G^{A}) &= 2 \, \partial_{[\mu} \omega_{\nu]}{}^{A} + 2 \, \omega_{[\mu|}{}^{A}{}_{B} \, \omega_{|\nu]}{}^{ B} +4 \, g_{[\mu} \, \tau_{\nu]}{}^{A} -4f_{[\mu}{}^{A} \, e_{\nu]}\,, \label{eq:RJ0A} \\
  R_{\mu\nu}(K) &= 2 \partial_{[\mu} g_{\nu]} + 2 \, \omega_{[\mu}{}^{A} \, f_{\nu]A} - 2 \, b_{[\mu}\, g_{\nu]} \,,  \\  
  R_{\mu\nu}(K^{A}) &= 2 \partial_{[\mu} g_{\nu]}{}^{A} + 2 \, \omega_{[\mu}{}^{AB} \, f_{\nu] B} - 2 \, b_{[\mu}\, f_{\nu]}{}^{A} \,,  \\
 R_{\mu\nu}(D) &= 2\, \partial_{[\mu} b_{\nu]} - 4\, f_{[\mu}{}^{A} \, \tau_{\nu]A} \,.
\end{align}
\end{subequations}
We will  make use of the dual vielbeine $(\tau_{A}{}^{\mu},e_{}^{\mu})$ defined through the following duality relations:
\begin{align} \label{invVielbdef}
    \tau_{A}{}^{\mu}\tau_\mu{}^{B} &= \delta_A^B\,, & \tau_A{}^{\mu}e_{\mu}{} &= 0\,, & e^{\mu}\tau_{\mu}{}^{A} &= 0\,,\notag \\ 
e_{\mu}{}e_{}^{\mu} &= 1\,, & e_{\mu}{}e^{\nu} &= \delta^{\nu}_{\mu}-\tau_{\mu}{}^{A}\tau_A{}^\nu\,.
\end{align}
We can see from \eqref{eq:confgaltrafos} that these dual vielbeine transform as follows:
\begin{align}
    \delta e^{\mu}&=\lambda_De^{\mu}\,,\notag \\
    \delta\tau_{A}{}^{\mu}&=-\lambda_A{}^{B}\tau_{B}{}^{\mu}-\lambda_Ae^{\mu}+\lambda_D \tau_A{}^{\mu}\,.
\end{align}
The multiplet of gauge fields obtained by gauging the conformal domain wall Galilean algebra contains more independent fields than are required for a gravitational theory. We therefore aim to impose suitable curvature constraints that allow one to express certain fields in terms of the remaining ones, thereby reducing the number of independent degrees of freedom.

Inspecting the curvatures listed in \eqref{eq:cgalconfcurvs}, one finds that a minimal set of independent fields can be achieved by constraining the curvatures $R_{\mu\nu}{}(P^{A})$ and $R_{\mu\nu}{}(P)$ associated with longitudinal and transversal translations, together with the components $R_{\mu B}{}(J^{AB})$ and $R_{\mu A}{}(G^{A})$ of the longitudinal rotations and boost curvatures. Consistent with the discussion in the preceding sections, one may impose the following constraints:
\begin{align} \label{eq:cgalconstr}
  R_{\mu\nu}(P) = \mathcal{E}_{\mu\nu} \,, \quad R_{\mu\nu}(P^A) = \mathcal{T}_{\mu\nu}{}^{A} \,, \qquad \qquad R_{\mu B}(J^{AB}) = 0 \,, \quad R_{\mu A}(G^{A}) = 0 \,,
\end{align}
where 
\begin{align}
    \mathcal{E}_{\mu\nu}&\equiv2\, \partial_{[\mu} e_{\nu]}{} + 2\, \omega_{[\mu}{}^{A} \, \tau_{\nu] A} +2b_{[\mu}e_{\nu]}= E_{\mu\nu}+2b_{[\mu}e_{\nu]}\,,\notag  \\
\mathcal{T}_{\mu\nu}{}^{A}&\equiv 2\, \partial_{[\mu} \tau_{\nu]}{}^{A} + 2\, \omega_{[\mu}{}^{AB} \, \tau_{\nu] B} + 2\, b_{[\mu} \, \tau_{\nu]}{}^{A}= {T}_{\mu\nu}{}^{A}+2\, b_{[\mu} \, \tau_{\nu]}{}^{A}\,. \label{TT}
\end{align}
Setting all non-intrinsic torsion components in \eqref{TT} equal to zero results in equations that allow to solve for the following dependent gauge field components:
\begin{align}
\omega_C{}^{AB}&= -\frac{1}{2}\tau^{AB}{}_{,C}+\tau_C^{[A,B]}+2b^{[A}\delta_C^{B]}\,,\notag\\
\omega_z{}^{AB}&= e^\nu{}\tau_\nu^{[A,B]}\,,\notag\\
\omega^{[A, B]}&=-\tau^{A\mu}\tau^{B\nu}\partial_{[\mu}e_{\nu]}\,,\notag \\
\omega_z{}^{A}&=-2e^{\mu}\tau^{A\nu}\partial_{[\mu}e_{\nu]}+\tau^{A\mu}b_\mu\,,\notag\\
b_{z} &=-\frac{2}{(D-1)} e^\mu \tau_A{}^\nu \partial_{[\mu} \tau_{\nu]}{}^A \equiv-\frac{1}{(D-1)}{T}_{z A,}{}^{A}\,,\label{wb}
\end{align}
with $\tau_{\mu\nu}{}^{A}\equiv2\partial_{[\mu}\tau_{\nu]}{}^{A}$. In an analogous manner, we can show that the last two constraints of \eqref{eq:cgalconstr} can be used to solve for the longitudinal and transversal special conformal gauge fields $f_\mu{}^{A}$ and $g_\mu $, as follows: 
\begin{equation} \label{fsol}
    f_{\mu}{}^{A}=-\frac{1}{2(D-3)}\bigg[R^\prime_{\mu B}\left(J^{AB}\right)-\frac{1}{2(D-2)}\tau_{\mu}{}^{A}R^\prime_{BC}{}\left(J^{BC}\right)-
    \frac{1}{(D-2)}e_{\mu}e^{\nu}R^\prime_{\nu B}\left(J^{AB}\right)\bigg]\,,
\end{equation}
\begin{equation} \label{fsol}
    g_{\mu}=-\frac{1}{2(D-2)}\left[R^\prime_{\mu A}\left(G^{A}\right)-\frac{e_{\mu}}{(D-1)}\left(e^{\nu}R^\prime_{\nu A}\left(G^{A}\right)+\frac{1}{2}R^\prime_{AB}\left(J^{AB}\right)\right)\right]\,.
\end{equation}
We then arrive at the following expressions, which will be useful in what follows \footnote{Note that the boost transformation of the combination \eqref{eq:specialcomb}, that will be crucial in what follows, is still given by the one that can be inferred from the transformation rules \eqref{eq:confgaltrafos}. This is however non-trivial since as dependent gauge fields, their transformations can pick up extra $\Delta$ contributions, as discussed in section \ref{ssec:pconfcarrgauging}. Indeed, both $\tau_A{}^\mu f_{\mu}{}^A$ and $e^\mu g_{\mu}$ acquire such extra terms in their boost transformations, but these cancel in their sum.}
\begin{eqnarray}
\tau_A{}^\mu f_\mu{}^A &=&-\frac{1}{4(D-2)}R^\prime_{AB}(J^{AB})\,,\\
e^\mu g_\mu &=& -\frac{1}{2(D-1)} e^{\nu}R^\prime_{\nu A}(G^{A}) + \frac{1}{4(D-1)(D-2)}R^\prime_{AB}(J^{AB})\,,\\
\tau_A{}^\mu f_\mu{}^A + e^\mu g_\mu &=& -\frac{1}{4(D-1)}\left[ R^\prime_{AB}(J^{AB}) + 2 e^{\nu}R^\prime_{\nu A}(G^{A})\right]\,. \label{eq:specialcomb}
\end{eqnarray}

\subsection{A conformal approach to Domain wall Galilean Gravity}
Let us now couple the conformal domain wall Galilean gauge fields constructed above to a massless scalar field. We will build two types of actions for $\phi$  which are invariant under homogeneous conformal transformations. As we shall see, upon gauge-fixing dilatations these actions reduce to domain wall Galilean gravity theories of the electric and magnetic types.\\
We assume that the compensating scalar field transforms under homogeneous conformal transformations as follows
\begin{equation}
    \delta \phi=w\lambda_D\phi\,.
\end{equation}
To build the conformal actions that are invariant under homogeneous conformal transformations, we define the covariant derivatives:
\begin{align}
 D_z\phi&\equiv e^{\mu}\left(\partial_\mu\phi-w b_{\mu}\phi\right)\,,\\
    D_A\phi&\equiv \tau_A{}^{\mu}\left(\partial_\mu\phi-w b_{\mu}\phi\right)\,,
\end{align}
whose transformation rule under homogeneous conformal transformations are given by
\begin{align}
 \delta D_z\phi&=(w+1)\lambda_DD_z\phi\,,\label{deltaDz}\\
\delta D_A\phi&=-\lambda_A{}^{B}D_B\phi+\lambda_{A} \, D_{z} \phi +(w +1)\lambda_D D_A\phi- 2 w \lambda_{K A} \, \phi \,.\label{deltaDA}
\end{align}
We will also require the following definition for the "transversal Laplacian" $D_zD_z\phi$:
\begin{align}
  D_{z}D_{z} \phi \equiv e^{\mu} \left[\partial_\mu\left(D_{z} \phi\right) - (w+1)\, b_\mu \, D_{z} \phi \right] \,,
\end{align}
that transforms under $\mathfrak{hcgal}_{\text{dw}}$ as follows,
\begin{align} 
\delta D_{z}D_{z} \phi = (w + 2) \, \lambda_D \,  D_{z}D_{z} \phi \,.
\end{align}
\vskip .1truecm

\subsubsection{Electric domain wall Galilei gravity.} 

We start by considering the following compensating transversal scalar:
\begin{equation}
    \mathcal{L}_{\text{transversal scalar}}=-\frac{1}{2}\phi\left(\partial_{z}\partial_{z}\phi\right)\,. \label{elecscal}
\end{equation}
We now promote the Lagrangian to be invariant under local homogeneous conformal domain wall  Galilean transformations. To this end, we first replace the time derivative of the scalar field by its covariant counterpart $D_z\phi$:
\begin{equation}
    \partial_{z}\phi\rightarrow D_z\phi\equiv e^{\mu}\left(\partial_\mu\phi-w b_{\mu}\phi\right)\,,
\end{equation}
whose transformation under homogeneous conformal transformations is given by \eqref{deltaDz}.
We consider the following Lagrangian describing the coupling of the transversal scalar to conformal domain wall Galilei gauge fields:
\begin{equation}
    \mathcal{L}_{\rm conformal\ coupling}=-\frac{1}{2} e\phi D_{z}D_{z} \phi\,. \label{Lag}
\end{equation}
Here, $e=\text{det}(e_{\mu},\tau_{\mu}{}^{A})$  which transforms under dilatations as
\begin{equation}\label{transfe}
\delta e = -D\lambda_D e\,.
\end{equation}
We find then that the Lagrangian \eqref{Lag} transforms as
\begin{equation}
   \delta\mathcal{L}_{\rm conformal\ coupling}=-\frac{1}{2}[-D+2w+2]\lambda_D\, e\phi D_{z}D_{z} \phi\,.
\end{equation}
This shows that the proposed Lagrangain is invariant under dilatations provided we take the scaling weight $w$ to be
\begin{equation}
     w=\frac{D-2}{2}\,.
\end{equation}
Having constructed the dilatation-invariant Lagrangian, we now fix the dilatations by setting $\phi=1$. This gauge choice reduces the theory to the following Lagrangian:
\begin{equation}
    \mathcal{L}_{\rm electric\ domain\ wall 
}= \frac{w^2}{2} e\, b_z^{2}\,.
\end{equation}
Substituting the expression for the dependent dilatation gauge field $b_z$ given in \eqref{wb}, we obtain a Lagrangian that describes a specific realization of electric domain wall gravity:
\begin{equation}
\mathcal{L}_{\rm electric\ domain\ wall}=\frac{w^2}{2(D-1)^{2}}\,e \,{T}_{ A}{}^{A}{T}_{ B}{}^{B}\,,
\end{equation}
where ${T}_{ z,A}{}^{A}\equiv{T}_{ A}{}^{A}$. Note that this is not the most general electric domain wall gravity theory given in the literature \cite{Henneaux:1979vn}. The following electric domain wall gravity was found in \cite{Bergshoeff:2023rkk}:
\begin{equation}
\mathcal{L}_{\rm electric\ domain\ wall} \sim \,-e \left(\, T_{}^{(A B)}T_{(A B)} - {T}_{ A}{}^{A}{T}_{ B}{}^{B}\right)\,.
\end{equation}
This invariant was obtained as an electric limit of general relativity in the second-order formulation. A noteworthy feature of this kind of electric invariants is that they are independent of the spin connection and therefore describe geometries without imposing any geometric constraints. In the absence of a spin connection, such invariants do not admit a first-order formulation.

\subsubsection{{Magnetic domain wall Galilei gravity}}
In this case, the action that we wish to make invariant under local homogeneous conformal Galilei transformations $\mathfrak{hcgal}_{\text{dw}}$ is that of a massless longitudinal scalar:
\begin{equation}
    \mathcal{L}_{\text{longitudinal scalar}}= -\frac{1}{2}\partial^A\phi\,\partial_A\phi+\chi\partial_z\phi\,, \label{longescalar}
\end{equation}
with $\chi$ an independent Lagrange multiplier field. As before, we extend the longitudinal and transversal derivatives $\partial_A\phi$ and $\partial_z\phi$ of $\phi$ to the following covariant derivatives:
\begin{align}
  \label{eq:DphiAM}
  D_{A} \phi&\equiv  \tau_{A}{}^{\mu} \left(\partial_\mu - w\, b_\mu \right) \phi \,,\notag\\
  D_{z}\phi &\equiv e_{}{}^{\mu} \left(\partial_\mu - w\, b_\mu \right) \phi \,,
\end{align}
whose transformation under $\mathfrak{hcgal}_{\text{dw}}$ is given by \eqref{deltaDz}-\eqref{deltaDA}. Then, we propose the following Lagrangian describing the coupling of the longitudinal scalar to conformal domain wall Galilei gauge fields:
\begin{equation}
    \mathcal{L}_{\rm Ansatz}= -\frac{1}{2}eD^{A}\phi D_{A} \phi+e\chi D_z \phi\,. \label{actlonggal}
\end{equation}
We proceed to check the invariance of the action under homogeneous conformal transformations. The invariance under longitudinal rotations is manifest from the construction. The variation of the action under a generic homogeneous conformal transformation is given by
\begin{eqnarray}
\delta\mathcal{L}_{\rm Ansatz}&=&-\frac{1}{2}\left( \delta e D^{A}\phi D_{A} \phi+2e D^{A}\phi \delta D_{A} \phi-2\delta e\chi D_z\phi-2e\delta \chi D_z \phi-2e\chi\delta D_z\phi\right)\notag\\
    &=&-\frac{1}{2}e\left[(-D+2(w+1))\lambda_DD^{A}\phi D_{A} \phi-2\lambda_A D^{A}\phi D_z\phi-2w\lambda_{K A}\phi D^{A}\phi\right.\notag \\
   &&\left.+2D\lambda_D \chi D_z\phi-2\delta \chi D_z \phi -2(w+1)\chi\lambda_DD_z \phi\right]\,.\label{deltaSdw2}
\end{eqnarray}
Then, the invariance of the Lagrangian under dilatations and domain wall Galilean boosts requires that 
\begin{equation}
    w=\frac{(D-2)}{2}\,,
\end{equation}
along with
\begin{equation}
\delta \chi=\lambda_A D^{A}\phi+\frac{1}{2}D\lambda_D\chi\,.
\end{equation}
From \eqref{deltaSdw2} we find that the action is not invariant under longitudinal special conformal transformations:
\begin{eqnarray}
\delta\mathcal{L}_{\rm Ansatz}
    &=&-\frac{1}{2}e\left(-2w\lambda_{K A}D^{A}\phi\right)\,.
\end{eqnarray}
This term can be canceled by adding two terms that involve the gauge fields of the longitudinal and transversal special conformal transformations. Then, we will consider the following action describing the coupling of a longitudinal scalar to conformal domain wall gauge fields:
\begin{equation}
    \mathcal{L}_{\rm conformal\ coupling}= -\frac{1}{2}eD^{A}\phi D_{A} \phi+e\chi D_z\phi+we f_A{}^{A}\phi^2+weg_z\phi^2\,,\label{confdwgalilei}
\end{equation}
where $g_z=e^\mu g_\mu$. With the addition of these terms, the action is now invariant under longitudinal rotations, boosts, dilatations and longitudinal special conformal transformations. One can show that invariance under transversal special conformal transformations is guaranteed allowing a $K$-variation of the Lagrange multiplier $\chi$, so that it transforms as follows:
\begin{equation}
\delta \chi=\lambda_A D^{A}\phi+\frac{1}{2}D\lambda_D\chi+(D-2)\lambda_K\phi\,.
\end{equation}
As in the previous cases, we fix the dilatations by setting $\phi=1$. Since the action \eqref{confdwgalilei} is invariant under longitudinal special conformal transformations and due to the fact that all dependent gauge fields transform under these transformations through their $b_A$-dependence the contributions involving $b_A$ drop out. The resulting expression defines the Lagrangian for magnetic domain wall Galilei gravity:
\begin{align}
    \mathcal{L}_{\text{magnetic domain wall}}&=w e\left(-\chi b_z+f_A{}^{A}+g_z\right)\,,\notag\\
    &=\frac{1}{2} \frac{D-2}{D-1}e\, \left\{\chi\, {T}_{A}{}^A
-\frac{1}{4}\left[ R^\prime_{AB}(J^{AB}) + 2 e^{\nu}R^\prime_{\nu A}(G^{A})\right]\right\}\,.\label{finaldwgal}
\end{align}
By gauge-fixing the $K$-transformations (for instance, setting $\chi=0$),
we are left with the following action describing domain wall Galilei gravity,
\begin{align}
    \mathcal{L}_{\text{magnetic domain wall}}
    &=-\frac{1}{8} \frac{D-2}{D-1} e\, 
\left[ R^\prime_{AB}(J^{AB}) + 2 e^{\nu}R^\prime_{\nu A}(G^{A})\right]\,.
\end{align}
This domain wall Galilean gravity action was obtained in \cite{Bergshoeff:2023rkk} as a limit of the Einstein-Hilbert action. 
\bibliography{draft}
\bibliographystyle{utphys}






\end{document}